\documentclass[12pt]{article}
\pdfoutput=1
\usepackage{a4wide}
\usepackage[centertags]{amsmath}

\usepackage{lmodern} % readable font in pdf
\usepackage{microtype} % better font spacing
\usepackage[english]{babel}

\usepackage{color,graphicx}
\usepackage[usenames,dvipsnames]{xcolor}

\usepackage{changes}

\usepackage{amsmath,dsfont,url,amssymb}
\usepackage{slashed,cancel}
\usepackage[usenames,dvipsnames]{xcolor} 
\usepackage[colorlinks=true, linkcolor=black, citecolor=ForestGreen]{hyperref}%BrickRed
\usepackage{mdframed}
\usepackage{subfigure,overpic}
\usepackage{feynmf}%%%{feynmp}
\usepackage[compat=1.1.0]{tikz-feynman}
\usepackage{textgreek}

\usepackage{graphicx}
\usepackage{sidecap}
\usepackage{float}
\usetikzlibrary{arrows.meta,calc}
\usepackage{xcolor}
\definecolor{arxivred}{HTML}{B31B1B}
\usepackage{xcolor}
\definecolor{arxivred}{HTML}{B31B1B}
\definecolor{deepblue}{HTML}{0B3C8C}
\definecolor{arxivred}{HTML}{B31B1B}
\definecolor{curveblue}{HTML}{2E6FCC} % lighter, more elegant blue
\usepackage{ragged2e}
\usepackage{microtype}
\usepackage{sidecap}
\usepackage{float}

\usepackage{amsmath}
\usetikzlibrary{arrows.meta,calc}

\title{%Boosted Hydrodynamic Convergence in Kinetic Theory and Holography
	Near-Light-Cone Nonhydrodynamic Structure from Boosted Hydrodynamics}
\author{
	Navid Abbasi$^{a,b,c}$,
	Jewel Kumar Ghosh$^{d,e}$,
	Dirk H.\ Rischke$^{b,f}$
	\\[2mm]
	\small{\textit{$^{a}$School of Nuclear Science and Technology, Lanzhou University, China}}\\
	\small{\textit{$^{b}$Institut f\"ur Theoretische Physik, Johann Wolfgang Goethe--Universit\"at, Germany}}\\
	\small{\textit{$^{c}$ExtreMe Matter Institute EMMI, GSI Helmholtzzentrum f\"ur Schwerionenforschung, Germany}}\\
	\small{\textit{$^{d}$Department of Physical Sciences, Independent University, Bangladesh (IUB), Bangladesh}}\\
	\small{\textit{$^{e}$Center for Computational and Data Sciences (CCDS), Independent University, Bangladesh}}\\
	\small{\textit{$^{f}$Helmholtz Research Academy Hesse for FAIR, Germany}}\\[1mm]
	\small{\texttt{abbasi@lzu.edu.cn}, \texttt{jewel.ghosh@iub.edu.bd}, \texttt{drischke@itp.uni-frankfurt.de}}
}

\begin{document}
	
	\setlength{\baselineskip}{16pt}
	\begin{titlepage}
		\maketitle
		
		\vspace{-36pt}
		
		%Abstract
		\begin{abstract}

			Hydrodynamics provides a universal description of many-body systems at long wavelengths. Its long-lived collective excitations, known as hydrodynamic modes, admit long-wavelength expansions whose convergence is encoded in the analytic structure of retarded correlators of conserved quantities in thermal equilibrium. We show that a boost can change which nonhydrodynamic singularity limits such an expansion. In the large-boost limit, the relevant singularity originates in the large-momentum, near-light-cone region of the rest-frame spectrum. After factoring out the Lorentz-factor growth, the rescaled convergence scale is controlled by the microscopic spectral structure of the theory. We explicitly demonstrate this in two examples: relaxation-time kinetic theory and holography.

		\end{abstract}
		\thispagestyle{empty}
		\setcounter{page}{0}
	\end{titlepage}

	\renewcommand{\baselinestretch}{1}  %looks better
	\tableofcontents
	\renewcommand{\baselinestretch}{1.2}  %looks better
	%%%%%%%%%%%%%%%%%%%%%%%%%%%%%%%%%%%%

	\maketitle
	%_____________________________________
	\section{Introduction}
	%_____________________________________
	Hydrodynamics is the effective description of long-time and long-wavelength dynamics in systems as diverse as relativistic plasmas, electron fluids, and cold atomic gases \cite{Florkowski:2017olj,Lucas:2017idv,Schaefer:2009dj}.
	Hydrodynamic gradient expansions in real space can be asymptotic
	\cite{Heller:2013fn}. The small-momentum expansion of a hydrodynamic mode, however, generally has a finite radius of convergence in the complex momentum plane \cite{Withers:2018srf,Grozdanov:2019kge,Heller:2020uuy}. As this expansion is continued away from $q=0$, the hydrodynamic mode eventually becomes sensitive to microscopic nonhydrodynamic degrees of freedom. The distance from the origin to the nearest such point in the complex momentum plane defines the radius of convergence $q_c$ of the hydrodynamic expansion. Beyond this radius, the hydrodynamic branch can be analytically continued using the information encoded in the spectral curve, i.e., the zeros of the inverse retarded correlator
	$F(\Omega,q)\equiv G_R^{-1}(\Omega,q)=0$ of a conserved quantity.
	
\begin{SCfigure}
		\centering
		\resizebox{0.52\columnwidth}{!}{
			%	\begin{tikzpicture}[
				%			>=Stealth,
				%			scale=1.0,
				%			axis/.style={black, line width=1.0pt, ->},
				%			curve/.style={red, line width=1.8pt},
				%			otherbranch/.style={red!85!black, dashed, line width=1.15pt},
				%			boost/.style={blue!85!black, line width=1.2pt, ->},
				%			boostdash/.style={blue!85!black, dashed, line width=1.0pt},
				%			nulldir/.style={gray!70!black, dashed, line width=1.0pt, ->},
				%			lab/.style={font=\large},
				%			smalllab/.style={font=\normalsize},
				%			pt/.style={circle, fill=black, inner sep=2.5pt},
				%			bluept/.style={circle, fill=blue!85!black, inner sep=2.9pt}
				%			]
				\begin{tikzpicture}[
					>=Stealth,
					scale=1.0,
					axis/.style={black, line width=1.0pt, ->},
					curve/.style={curveblue, line width=1.55pt, line cap=round},
					otherbranch/.style={curveblue!70!black, dashed, line width=1.05pt, line cap=round},
					boost/.style={arxivred, line width=1.15pt, ->},
					boostdash/.style={arxivred, dashed, line width=1.0pt},
					boostpt/.style={circle, fill=arxivred, inner sep=2.8pt},
					nulldir/.style={gray!65!black, dashed, line width=0.9pt, ->},
					lab/.style={font=\large},
					smalllab/.style={font=\normalsize},
					pt/.style={circle, fill=black, inner sep=2.5pt},
					boostpt/.style={circle, fill=arxivred, inner sep=2.9pt},
					boostgray/.style={gray!45, line width=0.8pt, ->},
					boostgraydash/.style={gray!45, dashed, line width=0.75pt}
					]
					
					% ============================================================
					% Panel (a): RF spectral curve and boosted projection
					% ============================================================
					
					\coordinate (O) at (0,0);
					
					% RF key points
					\coordinate (qc) at (2.70,1.56);
					\coordinate (qs) at (3.00,0.62);
					\coordinate (qs2) at (4.75,-1.47);
					
					% boosted image on k-axis
					\coordinate (kc) at (3.54,-0.76);
					\coordinate (kc2) at (5.9,-2.66);
					% axis endpoints
					\coordinate (Qend)  at (6.9,0);
					\coordinate (Omend) at (0,3.);
					\coordinate (Kend)  at (6.65,-1.35);
					\coordinate (wend)  at (-1.15,2.80);
					\coordinate (KendGray)  at (6.25,-2.85);
					\coordinate (wendGray)  at (-2.2,2.1);
					\coordinate (nend)  at (3.55,-2.50);
					
					% panel title
					\node[smalllab, anchor=west] at (-1.95,3.6)
					{(a) 
						%RF spectral curve and boosted projection
					};
					
					% RF axes
					\draw[axis] (-0.20,0) -- (Qend) node[right] {$q$};
					\draw[axis] (0,-0.25) -- (Omend) node[above] {$\Omega$};

					% second boosted frame, larger boost, shown in gray
					%	\coordinate (KendGray) at (5.75,-2.05);
					%	\coordinate (wendGray) at (-2.05,2.55);
					
					\draw[boostgray] (O) -- (KendGray)
					node[right, gray!70!black, font=\normalsize] {$k'$};
					
					\draw[boostgray] (O) -- (wendGray)
					node[above left, gray!70!black, font=\normalsize] {$\omega'$};

					% boosted axes
					\draw[boost] (O) -- (Kend) node[right] {$k$};
					\draw[boost] (O) -- (wend) node[above left] {$\omega$};

					% null direction
					%	\draw[nulldir] (O) -- (nend)
					%	node[midway,right=1pt] {\small null direction};
					
					% auxiliary dashed boosted-omega direction
					\draw[boostdash] (O) -- (0.765,-1.8);
					\draw[boostdash,gray] (O) -- (2.45,-2.3);
					
					% main RF branch
					\draw[curve]
					(O)
					.. controls (0.10,0.02) and (0.48,0.04) .. (1.18,1.20)
					.. controls (1.78,2.15) and (2.41,2.66) .. (qc)
					.. controls (2.82,0.82) and (2.98,0.70) .. (qs)
					.. controls (3.12,0.15) and (4.55,-1.52) .. (6.05,-2.45);
					
					% added branch from qc
					\draw[otherbranch]
					(qc)
					.. controls (3.15,1.92) and (4.55,2.45) .. (6.05,3.05)
					node[pos=0.54, above, sloped, black, font=\small] {other branch};
					
					% qc
					\node[pt] at (qc) {};
					\draw[black, dashed] (2.70,0) -- (qc);
					\node[lab, anchor=south west] at ($(qc)+(-0.56,-.05)$) {$q_c$};
					
					% q*(v)
					\node[pt, arxivred] at (qs) {};
					\node[lab, arxivred, anchor=west] at ($(qs)+(0.18,-0.02)$) {$q_*(v)$};
					
					% tangent / projection to kc(v)
					\draw[boostdash] ($(qs)+(0,-0.05)$) -- (kc);
					
					% q2*(v)
					\node[pt, gray] at (qs2) {};
					\node[lab, gray, anchor=west] at ($(qs2)+(0.18,-0.02)$) {{$q_*(v')$}};
					
					% tangent / projection to kc(v)
					\draw[boostdash,gray] ($(qs2)+(0,-0.05)$) -- (kc2);

					%\node[boostpt] at (kc) {};
					\node[lab, arxivred, anchor=north] at ($(kc)+(0,0.0)$) {$k_c(v)$};
					\node[lab, gray, anchor=north] at ($(kc2)+(-.5,0.05)$) {$k_c(v')$};
					
					%\draw[black, dashed]  (qc)+(0.1,0) -- (3.76,-0.75);

					% ============================================================
					% Panel (b): boosted view omega(k)
					% ============================================================

					\begin{scope}[shift={(0,-7.550)}]
						
						\coordinate (BO) at (0,0);
						\coordinate (BKend) at (6.55,0);
						\coordinate (BWend) at (0,3.5);
						
						% branch-point image kc(v)
						\coordinate (Bkc) at (3.35,.75);
						\coordinate (Bkcaxis) at (3.35,0);

						% image of qc on upper branch
						\coordinate (Bqc) at (3.68,1.88);
						
						% panel title
						\node[smalllab, anchor=west] at (-1.95,3.2)
						{(b) 
							%boosted view: $\omega(k)$
						};
						
						% blue LAB axes
						\draw[boost] (-0.20,0) -- (BKend) node[right] {$k$};
						\draw[boost] (0,-0.25) -- (BWend) node[above] {$\omega$};
						
						% single-valued branch up to kc(v)
						\draw[curve]
						(BO)
						.. controls (0.65,0.25) and (1.2,.75) .. (1.8,1.45)
						.. controls (2.2,1.98) and (2.5,2.35) .. (2.58,2.45)
						.. controls (3.1,3.15) and (3.9,2.80) .. (Bqc)
						.. controls (3.6,1.55) and (3.55,1.3) .. (3.5,1.22)
						.. controls (3.5,1.15) and (3.25,.73) .. (Bkc);
						% lower branch after kc(v)
						\draw[curve]
						(Bkc)
						.. controls (3.3,0.4) and (3.6,.0) .. (5.10,-1.);

						% upper loop/branch after kc(v), containing image of qc
						%	\draw[curve]
						%	(Bkc)
						%					.. controls  (4.05,2.18) and (4.25, 2.16).. (4.45,2.12)
						%	.. controls (4.78,2.03) and (4.55,1.68) .. (Bqc);
						
						% continuation from image of qc to upper-right
						\draw[otherbranch]
						(Bqc)
						.. controls (4.55,1.82) and (5.15,2.45) .. (6.25,3.15);
						
						% point kc(v)
						\node[boostpt] at (Bkc) {};
						\draw[black, dashed] (Bkcaxis) -- (Bkc);
						\node[lab, arxivred, anchor=north] at ($(Bkc)+(0.05,-0.82)$) {$k_c(v)$};
						
						% point qc image
						\node[pt] at (Bqc) {};
						%\node[lab, anchor=west] at ($(Bqc)+(0.10,-.15)$) {$q_c$};
						
					\end{scope}
					
				\end{tikzpicture}
			}
			\vspace{8ex}
			\caption{
				(a) Schematic RF spectral curve (blue). The hydrodynamic branch passes through the origin. RF variables correspond to the black axes. Red axes indicate boosted variables for velocity $v$, while gray axes illustrate a larger boost $v'>v$. The point $q_c$ corresponds to the convergence radius of the hydrodynamical expansion in the RF, while $q_\ast(v)$ is the critical point satisfying $dk/dq=0$ and determines the convergence scale $k_c(v)=|k(q_\ast(v);v)|$ in the boosted frame.
				Note that a larger boost moves that point further along the analytically continued RF branch.
				(b) Spectral curve in a frame moving with $v$. 
				The black point is the boosted image of $q_c$ and need not determine the boosted convergence radius.  
			}
			\vspace{-2ex}
			\label{fig:boost_geometry}
		\end{SCfigure}
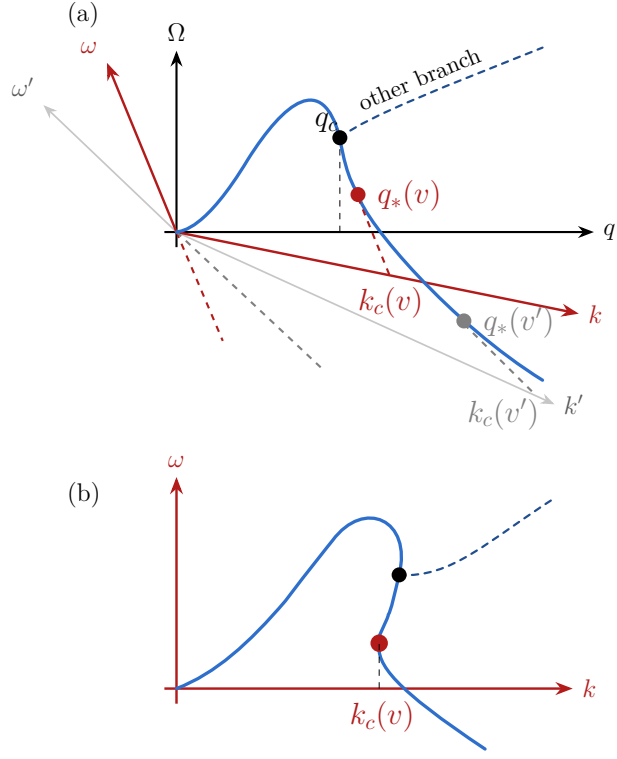

	In relativistic systems this raises a natural question. 
	\textit{If the same fluid is viewed from a frame moving with velocity $v$, is the breakdown scale of hydrodynamics given by  $q_c$ in the rest frame (RF) when represented in boosted variables?} 
	Since the hydrodynamic equations and their dispersion relations transform covariantly, such an expectation seems natural. 
	We show, however, that this is in general incorrect. 
	As illustrated in Fig.~\ref{fig:boost_geometry}, the boost reparametrizes the entire RF spectral curve by the boosted momentum $k$. 
	Note that, in this boosted frame, analyticity is already lost at the point $k_c(v)$.
	On the RF curve, this point, which we denote by $q_*(v)$ and whose selection is determined below, actually corresponds to a different point than $q_c$. 
	This observation motivates the central question of this Letter: \textit{which part of the RF spectral curve is selected by the
		boosted convergence scale \(k_c(v)\)?}

	%%%%%%%%%%%%%%
	%\begin{equation}\label{map}
	%	k(q; v)=\gamma\bigl(q+v\Omega(q)\bigr).
	%\end{equation}
	%%%%%%%%%%%%%%

		For a longitudinal boost, the RF hydrodynamic branch $\Omega=\Omega(q)$ is represented in boosted variables as the parametric curve $\bigl(\omega(q;v),k(q;v)\bigr).$ 
		Although this curve transforms covariantly, the boosted hydrodynamic dispersion relation is the function $\omega(k)$, obtained only after eliminating the RF momentum $q$.  
		This step is analytic only when the map $q\mapsto k(q;v)$ is locally invertible.  
		It fails when
		%%%%%%%%%%%%%%%%%
		\begin{equation}\label{loss}
			J(q;v)\equiv \frac{dk}{dq}
			=\gamma\left[1+v\Omega'(q)\right]=0 \;.
		\end{equation}
		%%%%%%%%%%%%%%%%%
		Indeed, near a generic zero of $dk/dq$, one has $k-k_*\sim (q-q_*)^2$ while $\omega-\omega_*\sim q-q_*$. 
		Thus, when the same curve is viewed as a function $\omega(k)$, it develops the local square-root form $\omega(k)-\omega_* \sim \sqrt{k-k_*}$.
		The boosted hydrodynamic dispersion relation therefore has a branch point that does not need to coincide with the boosted representation of the RF convergence-limiting point $q_c$ (cf.~lower panel).
		In fact, the point $q_*(v)$ may lie beyond $q_c$ on the RF hydrodynamic branch analytically continued to complex momentum (cf.~upper panel). 
		Since the projection $q\mapsto k(q;v)$ becomes locally non-invertible at $q_*(v)$, the image $k(q_*(v);v)$ can be closer to the origin in the boosted $k$-plane than the boosted image of $q_c$. 
		
		In the large-boost limit, Eq.~\eqref{loss} gives
		%%%%%%%%%%%%%%%%%s s
		\begin{equation}\label{large}
			\Omega'(q_*)=-\frac{1}{v}\to -1 \;.
		\end{equation}
		Thus the selected point is driven to a region of the RF spectral curve where
		the local slope approaches the light-cone value. If $q_\ast(v)$ had a finite limit as $v\to1$, this would require an exceptional finite-momentum point on the analytically continued hydrodynamic branch with $\Omega'(q)=-1$. 
		In the examples studied below, no such finite-momentum obstruction is encountered. 
		Instead $q_\ast(v)$ moves to infinite $|q|$, where the relevant RF nonhydrodynamic
		branch approaches the light-cone,
		%%%%%%%%%%%%%%%%%
		\begin{equation}\label{}
			\Omega(q) \sim -q+\delta(q)\;,\qquad \frac{|\delta(q)|}{|q|}\to 0 \;.
		\end{equation}
		This is the sense in which the large-boost convergence-limiting point $q_*(v)$ probes the
		near-light-cone part of the RF spectrum.

		What remains is a dynamical question: \textit{how does the microscopic spectral curve approach the light-cone, and what scale, if any, controls its offset $\delta_\infty \equiv \lim_{q\rightarrow \infty} \delta (q)$ from it?} 
		Since this is not fixed by Lorentz covariance or hydrodynamics \footnote{Related constraints on boosted relaxation spectra and the validity of hydrodynamics in boosted frames were derived for Onsager-symmetric and purely relaxational theories in Refs.~\cite{Gavassino:2026klp,Gavassino:2026seq}.}, we study this mechanism in two stable and causal settings: quasiparticle kinetic theory in relaxation-time approximation (RTA) and standard Einstein-Maxwell holography with second-order derivatives.
		The resummed diffusion equation derived from the former case provides an analytically tractable example with a finite microscopic relaxation rate. 
		Holographic charge diffusion provides a strongly coupled example, in which the relevant nonhydrodynamic excitations are quasi-normal modes (QNM) rather than quasiparticle relaxation modes.
		
		Hydrodynamic fluctuation effects, such as branch cuts and long-time tails generated by convolution of hydrodynamic modes~\cite{Kovtun:2003vj}, are parametrically suppressed in the limits considered here: the classical large-$N$ limit in holography~\cite{CaronHuot:2009iq} and the kinetic large-occupancy regime of the RTA-derived resummed diffusion theory~\cite{Chen-Lin:2018kfl}.
		
		The two examples exhibit distinct near-light-cone behaviors. 
		In RTA, the convergence-limiting point $q_*(v)$ is driven to large RF momentum and approaches the light-cone direction with a finite offset $\delta_\infty$ set by the relaxation time $\tau$, so that $k_c(v)/\gamma$ approaches a nonzero limit. 
		In holography, the convergence is limited by a QNM collision at a complex momentum, which under boosts is driven toward the light-cone itself, and $k_c(v)/\gamma\to0$. 
		Thus the boost dependence of the hydrodynamic radius of convergence is not merely kinematic; it diagnoses how microscopic nonhydrodynamic physics approaches the light-cone.

		%___________________________________________________________________
		\section{Kinetic theory in RTA and the boosted diffusion sector}
		\label{RTAsetup}
		%___________________________________________________________________
		In RTA-derived resummed diffusion theory the spectral curve in a frame boosted with velocity \(v\) is determined from 
		%%%%%%%%%%%%%%%%%%
		\begin{equation}\label{mode_resummed_boost_frame}
			F(\omega, k;v)\equiv 	1-\frac{1}{2 i \tau Q}\ln \Bigg[\frac{1+ i \tau Q- i \tau\gamma\big(\omega-\textbf{v} \cdot \textbf{k} \big) }{1- i \tau Q- i \tau\gamma\big(\omega-\textbf{v} \cdot \textbf{k} \big) }\Bigg]=0\;,
		\end{equation}
		%%%%%%%%%%%%%%%%%%
		where $Q=\sqrt{\Big[\gamma \omega \boldsymbol{v} - \textbf{k}-(\gamma-1)\hat{\textbf{v}} (\hat{\textbf{v}} \cdot \textbf{k})\Big]^2}$. 
		The above equation reduces to the familiar result (found in Refs.~\cite{Romatschke:2015gic,Kurkela:2017xis,Abbasi:2024pwz}) in the RF.
		In the RF, the solution of Eq.~\eqref{mode_resummed_boost_frame}  corresponding to the hydrodynamic branch is 
		%%%%%%%%%%%%%%%%%%%
		\begin{equation}\label{branch}
			\Omega(q) \;=\; -\frac{i}{\tau} + i q \cot(\tau q)\;.
		\end{equation}
		%%%%%%%%%%%%%%%%%%%
		
		Considering momenta parallel to the boost direction, \(\mathbf k\parallel \mathbf v\), the boosted image of the RF branch is then
		%%%%%%%%%%%%%%%%%%%
		\begin{equation}		\label{boosted_parametric_branch}
			k(q;v)=\gamma\bigl[q+v\,\Omega(q)\bigr]\;,\qquad
			\omega(q;v)=\gamma\bigl[\Omega(q)+v q\bigr]\;.
		\end{equation}
		%%%%%%%%%%%%%%%%%%%
		The convergence-limiting points of the boosted dispersion relation are therefore obtained from the extrema of the map \(q\mapsto k(q;v)\), determined by Eq.~\eqref{loss}.
		For each solution \(q_*(v)\), the corresponding singular point in the boosted variables is
		\begin{equation}
			k_*(v)=k(q_*(v);v)\;,\qquad
			\omega_*(v)=\omega(q_*(v);v)\;,
			\label{eq:RTA-critical-point}
		\end{equation}
		and the radius of convergence is the smallest such critical value,
		\begin{equation}
			k_c(v)=\min_{q_*}|k_*(v)|\;.
			\label{eq:RTA-radius}
		\end{equation}
		
		At \(v=0\), the RF branch~\eqref{branch} has nonanalyticities at the poles of \(\cot(\tau q)\),
		\begin{equation}
			\tau q=n\pi\;,\qquad n\in\mathbb Z \;,
			\label{eq:RTA-poles}
		\end{equation}
		except for $n=0$, where $\Omega(q=0) =0$.
		For \(v\neq0\), however, these poles are mapped to \(|k|=\infty\) and do not give finite convergence-limiting points of the boosted dispersion relation. 
		The convergence-limiting points are instead the points $q_*(v)$ determined by Eq.~\eqref{loss}.  
		We follow these solutions by continuation in \(v\), initializing near the \(n\)-th pole $\tau q=n \pi$ of $\cot(\tau q)$ at small boost and using the solution at the previous value of \(v\) as the initial guess.

		%%%%%%%%%%%%%%%%%%%%%%%%
		\begin{SCfigure}
			\centering
			\includegraphics[width=.45\textwidth]{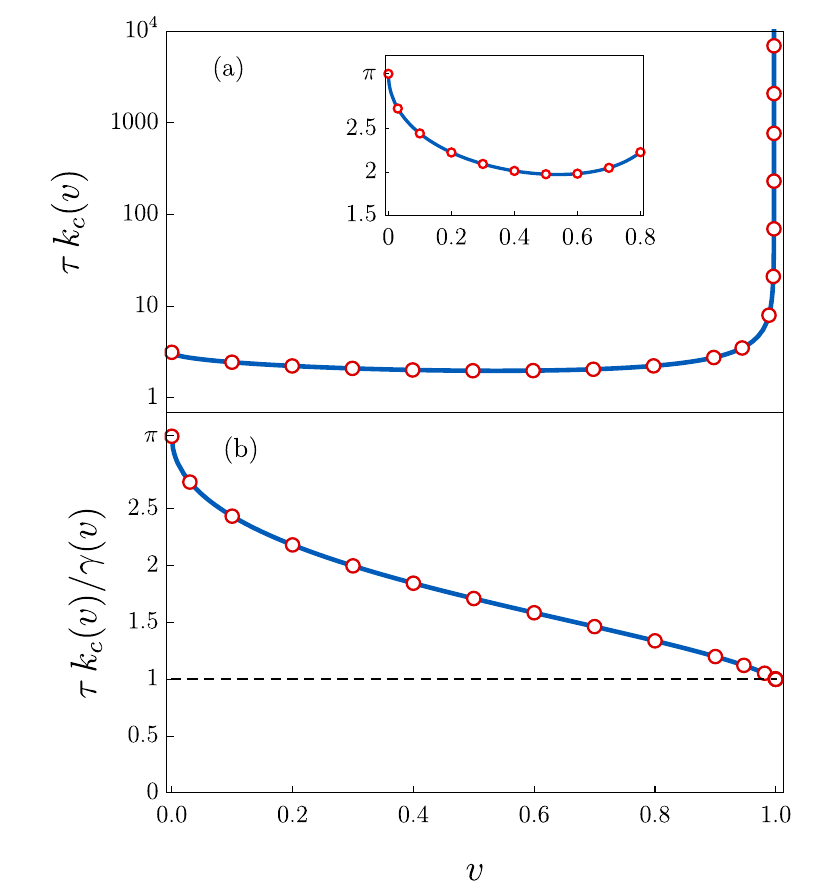}
			\caption{Critical momentum in the boosted RTA diffusion channel.
				(a) The radius of convergence \(k_c(v)\) grows rapidly as \(v\to 1\);
				the inset shows the same data on a linear scale away from the endpoint. Red circles are numerical solutions of the critical-point equation~\eqref{loss} at the sampled boost velocities; the blue line is an interpolation to guide the eye.
				(b) After rescaling by the Lorentz factor, \(k_c(v)/\gamma(v)\) approaches
				a finite value, \(\tau k_c(v)/\gamma(v)\to 1\), in the ultra-relativistic limit. }
			\label{RTA_radius}
		\end{SCfigure}
		%%%%%%%%%%%%%%%%%%%%%%%%

		The result is presented in the top panel of Fig.~\ref{RTA_radius}, which shows that \(\tau k_c(v)\) remains of order unity at moderate $v$, but grows rapidly as \(v\to1\).
		The more revealing quantity is the radius rescaled by the Lorentz gamma factor shown in the bottom panel: \(\tau k_c(v)/\gamma(v)\) decreases monotonically from \(\pi\) and approaches unity in the ultra-relativistic limit.
		%%%%%%%%%%%%%%%%%%%%%%%%

		We now show analytically why this limiting value is fixed by the relaxation rate $\tau$.
		Let $z=\tau q$, $s=e^{-2iz}$, and $\epsilon=v^{-1}-1 >0$. 
		Then, for the RF branch \eqref{branch}, the critical-point condition~\eqref{loss} can be written as
		\begin{equation}
			\Omega'(q_\ast)+1
			=
			{s_\ast(4iz_\ast-2+2s_\ast)\over(1-s_\ast)^2}
			=-\epsilon \;.
		\end{equation}
		As $v\to1$, $\epsilon\to0$. 
		In this limit, we either have to require $s_*\to 0$ or \(4iz_*-2+2e^{-2iz_*} \to 0\). 
		The second condition can only be fulfilled for $z_* \to 0$, which implies $q_* \to 0$, which is not a critical point on the hydrodynamic RF branch (see Appendix~\ref{large_limit}). 
		Hence \(s_\ast\to0\), or equivalently ${\rm Im}\,z_\ast\to-\infty$. 
		In this limit, the RF branch becomes
		\begin{equation}
			\Omega(q)=-q-{i\over\tau}+O(qe^{-2i\tau q})\;,
			\label{approach_RTA}
		\end{equation}
		and therefore
		\begin{equation}
			{k_\ast(v)\over\gamma(v)} = q_* + v\Omega(q_*)
			\longrightarrow -{i\over\tau}\;.
		\end{equation}
		Thus,
		\begin{equation}
			\frac{\tau k_c(v)}{\gamma(v)}\longrightarrow 1 \;.
			\label{eq:RTA-radius-limit}
		\end{equation}
		Thus the limiting singularity in RTA approaches the RF light-cone but remains a fixed distance away from it, with offset  given by the microscopic relaxation rate \(1/\tau\).

		This limit has a simple kinetic interpretation and confirms the geometric picture described in the Introduction.  
		Namely, in the absence of collisions, massless quasiparticles propagate freely along the light-cone, i.e., \(\Omega = - q\).
		The finite RTA relaxation time dampens this ballistic signal as \(e^{-t/\tau}\).  
		In frequency space this does not change the light-like slope, but shifts the dispersion relation of this signal away from the light-cone by the collision rate.

		%____________________________________________________________-
		\section{Holography and boosted Maxwell diffusion}
		%____________________________________________________________-
		We now turn to a strongly coupled  scenario. 
		Consider a probe \(U(1)\) current in the AdS\(_5\)-Schwarzschild black-brane background.  
		The diffusion pole and its nonhydrodynamic partners are obtained from the Maxwell QNM problem for gauge-invariant master fields \cite{Kovtun:2005ev} \footnote{Related holographic constructions of all-order \(U(1)\) diffusion and
			momentum-dependent transport functions were developed in
			Ref.~\cite{Bu:2015ame}.}.  
		Our goal is to track how the QNM spectrum in the longitudinal channel reorganizes under a longitudinal boost and to identify the nearest limiting singularity to analytic continuation in complex momentum, thereby defining the holographic analogue of \(k_c(v)\). 
		We use the package of Ref.~\cite{Jansen:2017oag} to find the QNM spectrum.
		
		Taking the radial coordinate to be \(u\), with boundary at \(u=0\) and horizon at \(u=1\), we write the Maxwell perturbation as \(A_M(u,x^\mu)=a_M(u)e^{-i\Omega t+iqz}\).  
		In the longitudinal channel, the gauge-invariant combination
		\begin{equation}
			E_\parallel = q a_t+\Omega a_z
		\end{equation}
		obeys
		\begin{equation}
			E_\parallel''(u)
			+\frac{\Omega^2 f'(u)}{f(u)\left[\Omega^2-q^2 f(u)\right]}
			E_\parallel'(u)
			+\frac{\Omega^2-q^2 f(u)}{u f(u)^2}E_\parallel(u)=0\;,
			\label{eq:maxwell_master}
		\end{equation}
		where \(f(u)=1-u^2\).  
		The QNMs are the values of \((\Omega,q)\) for which Eq.~\eqref{eq:maxwell_master} admits a nontrivial solution satisfying ingoing boundary conditions at the horizon and Dirichlet boundary conditions at the AdS boundary.

		The unboosted master equation \eqref{eq:maxwell_master} is naturally written in terms of the Lorentz scalars
		\[
		\Omega=-U^\mu K_\mu\;,\quad
		q^2=\Delta^{\mu\nu}K_\mu K_\nu\;,\quad
		\Delta^{\mu\nu}=\eta^{\mu\nu}+U^\mu U^\nu \;,
		\]
		where \(U^\mu\) is the fluid four-velocity and \(K^\mu=(\omega,\mathbf k)\) is the boosted-frame four-momentum.
		Thus a boost does not change the equation itself; it changes which RF values \((\Omega,q)\) are sampled by a given boosted-frame pair \((\omega,k)\), as given by Eq.~\eqref{boosted_parametric_branch}.
		
\begin{figure}[t]
			\centering
			\vspace{-1.0ex}
			
			\begin{minipage}[t]{0.315\textwidth}
				\centering
				\begin{overpic}[width=\linewidth]{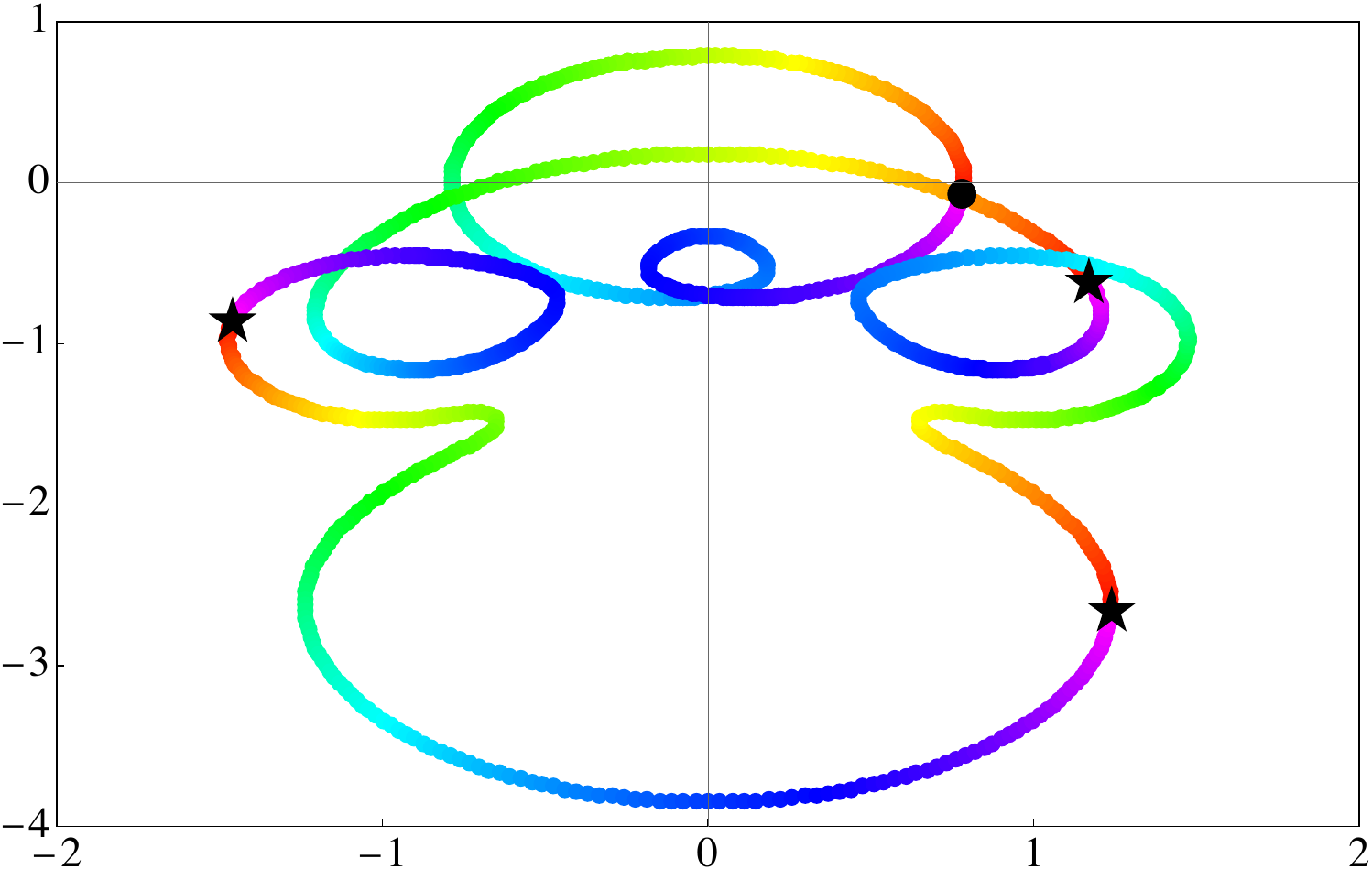}
					\put(47,-5){\small $\mathrm{Re}\,\frac{\omega}{T}$}
					\put(-5,30){\rotatebox{90}{\small $\mathrm{Im}\,\frac{\omega}{T}$}}
					\put(6,68){\small (a)}
				\end{overpic}
			\end{minipage}
			\hfill
			\begin{minipage}[t]{0.315\textwidth}
				\centering
				\begin{overpic}[width=\linewidth]{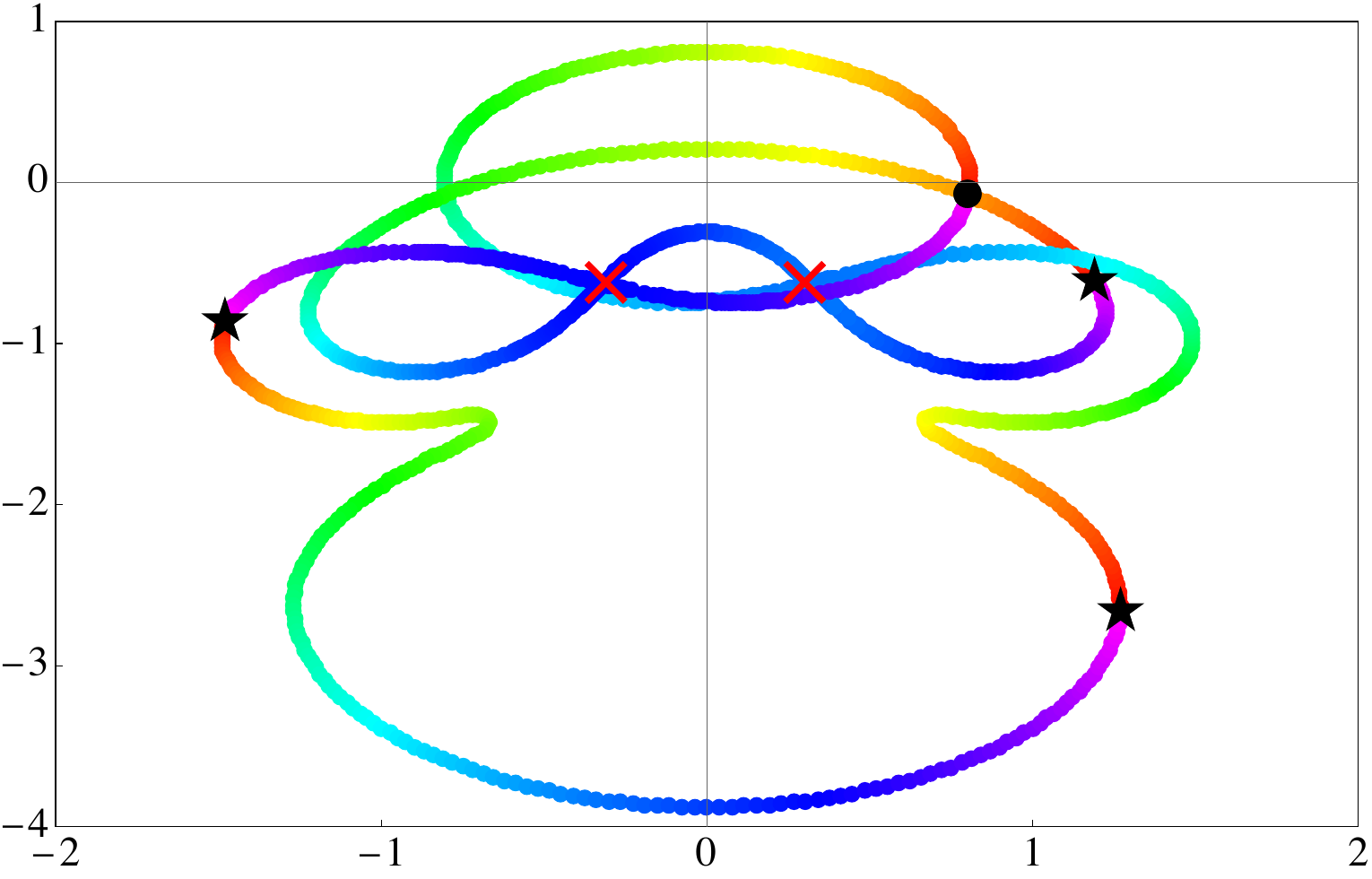}
					\put(47,-5){\small $\mathrm{Re}\,\frac{\omega}{T}$}
					\put(-5,30){\rotatebox{90}{\small $\mathrm{Im}\,\frac{\omega}{T}$}}
					\put(6,68){\small (b)}
				\end{overpic}
			\end{minipage}
			\hfill
			\begin{minipage}[t]{0.315\textwidth}
				\centering
				\begin{overpic}[width=\linewidth]{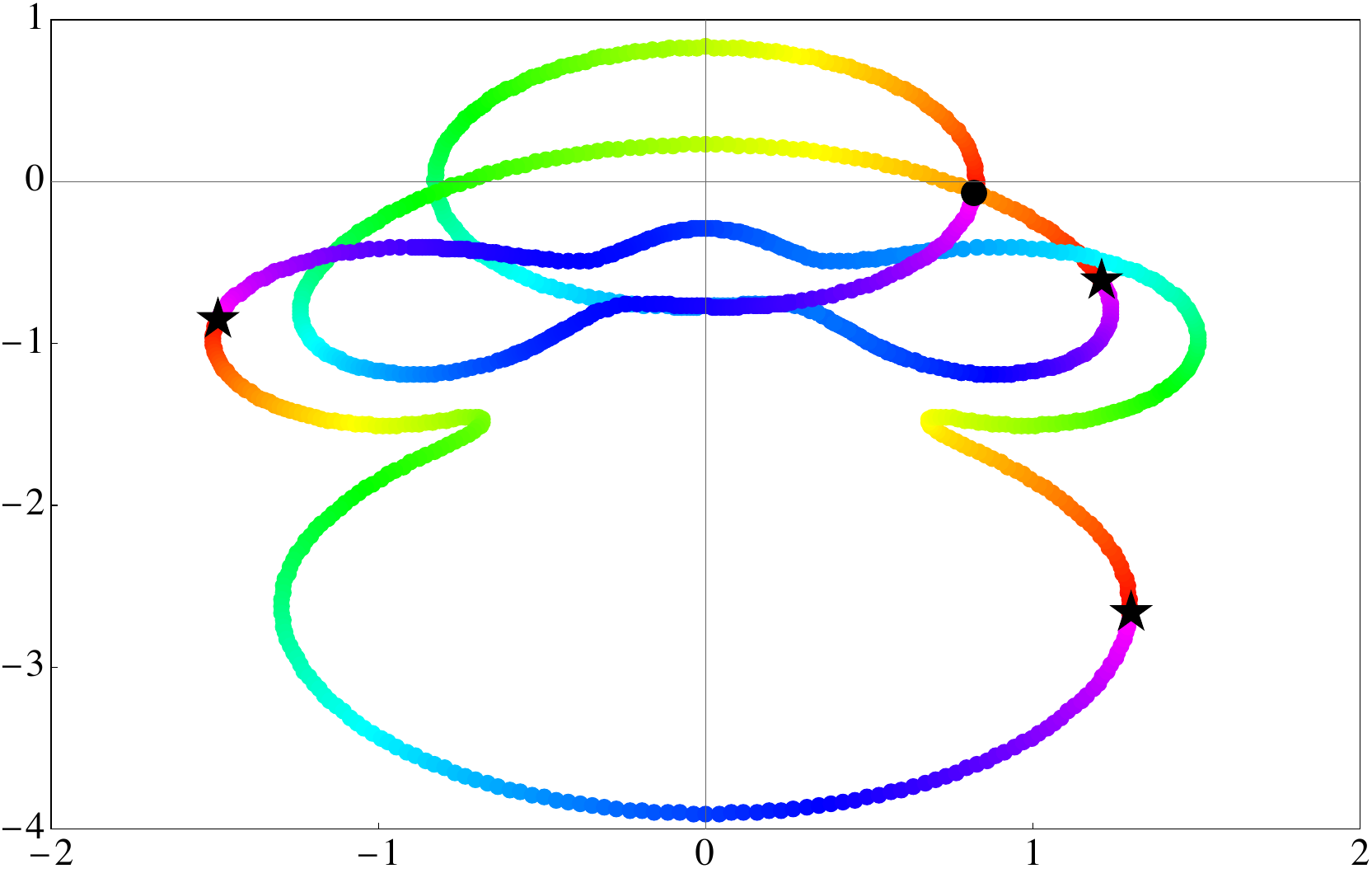}
					\put(47,-5){\small $\mathrm{Re}\,\frac{\omega}{T}$}
					\put(-5,30){\rotatebox{90}{\small $\mathrm{Im}\,\frac{\omega}{T}$}}
					\put(6,68){\small (c)}
				\end{overpic}
			\end{minipage}
			
			\vspace{-0.5ex}
			\caption{
				Analytic continuation of the longitudinal QNM spectrum at
				\(v=0.85\). The boosted QNM momentum is analytically continued as \(k=|k|e^{i\theta}\). The black dot corresponds to the hydrodynamic mode at real momentum, $\theta =0$. Black stars denote nonhydrodynamic modes at real momentum, \(\theta=0\). The colored trajectories (from red to magenta) show the motion of the modes in the complex \(\omega\)-plane as \(\theta\) is varied from 0 to $2 \pi$. The panels are for (a) \(|k|=k_c-\delta k\),
				(b) \(|k|=k_c\), and (c) \(|k|=k_c+\delta k\), with $\delta k= 0.01\,T$.  
				The red markers in the middle panel show the collision between the hydrodynamic and the nearest nonhydrodynamic mode, identifying the branch point that sets the
				radius of convergence at this boost velocity.}
			\label{fig:trajectory}
			\vspace{-1.0ex}
		\end{figure}
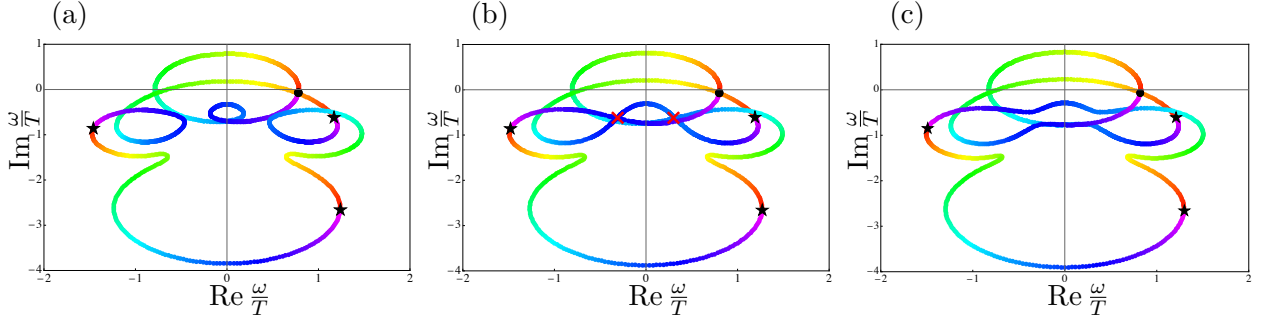

		The covariant form also makes clear why the boosted spectrum differs qualitatively from the familiar RF Maxwell spectrum of Ref.~\cite{Kovtun:2005ev}.  
		At \(v=0\), the QNM pattern is symmetric in the complex-frequency plane.  
		A longitudinal boost breaks this symmetry in the boosted-frame variables \((\omega,k)\), since the RF frequency and momentum are sampled through the combinations \(\Omega=\gamma(\omega-vk)\) and \(q=\gamma(k-v\omega)\). 
		This effect is clearly visible in Fig.~\ref{fig:trajectory}: at real $k$, the QNMs, indicated by black markers, are no longer symmetric about the imaginary axis in the boosted-frame frequency plane.

		%%%%%%%%%%%%%%%%%%%%%%%%%%%%%%%%%%
		%%%%%%%%%%%%%%%%%%%%%%%%%%%%%%%%%%			

		The convergence problem, however, requires more than the real-$k$ spectrum. 
		When the hydrodynamic mode is continued to complex boosted momentum, two limiting points are found, which are shown as red markers in the middle panel of Fig.~\ref{fig:trajectory}. 
		Their modulus gives the convergence radius at this boost velocity. 
		The noteworthy feature is that there are actually two such limiting points and that they are symmetric under reflection about the imaginary-$\omega$ axis, even though the real positive-$k$ spectrum shown by the black markers is not.
		This follows from the real-valuedness of the retarded correlator in coordinate space, which implies  $G_{R}^*(\omega,k)=G_R(-\omega,-k)$ in momentum space.
		Equivalently, the spectral equation satisfies $F(\omega,k;v)= F(-\omega^*,-k^*;v)=0$.
		This holds also for the critical point $(\omega_*(v),k_*(v))$, which implies that, in the QNM spectrum analytically continued to complex momenta, there must be two such points which are symmetric under reflection about the imaginary-$\omega$ axis.
		%Indeed, under $(\omega,k)\mapsto(-\omega^*,-k^*),$ the RF variables transform as $\Omega=\gamma(\omega-vk)\mapsto-\Omega^*,$ and $q=\gamma(k-v\omega)\mapsto-q^*$, which is the same symmetry in the RF variables. 
		%For real positive $k$, the partner lies at negative real momentum and is therefore not among the black markers shown at $\theta=0$. 
		%But along the analytic continuation $k=|k|e^{i\theta}$, both partners lie on the same $|k|=k_c$ circle. 
		%Their images in the $\omega$-plane are thus reflected about the imaginary axis, explaining the pair of symmetric red markers in Fig.~\ref{fig:trajectory}(b).
		
		Having illustrated the analytic continuation at one representative boost, we now extract the convergence radius as a function of \(v\). 
		Unlike in the resummed RTA-diffusion theory, we do not have a closed-form spectral equation the branches of which can be followed analytically. 
		Instead, we locate the limiting singularity numerically. 
		For each value of \(v\), we continue the hydrodynamic mode to complex boosted momentum and scan over the relevant nonhydrodynamic QNM tower. 
		The first limiting singularity (such as the collision points in Fig.~\ref{fig:trajectory}(b)) encountered in this continuation determines \(k_c(v)\).
		
		\begin{SCfigure}
			\centering
			\includegraphics[width=.45\textwidth]{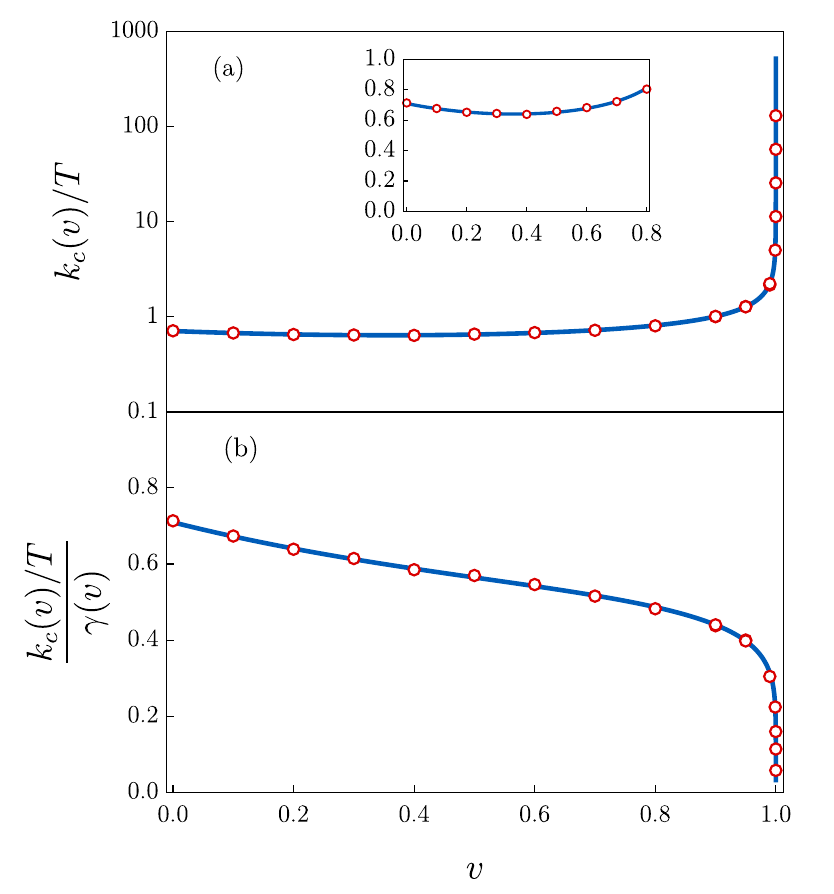}
			\caption{
				Critical momentum in the boosted holographic $U(1)$ diffusion channel.
				(a) The convergence radius $k_c(v)/T$ grows rapidly as $v\to1$; the inset
				resolves the moderate-boost region on a linear scale.  (b) After rescaling
				by the Lorentz factor, $(k_c/T)/\gamma(v)$ decreases toward zero, in sharp
				contrast to the RTA result of Fig.~\ref{RTA_radius}.  The solid curves are
				smooth fits to the numerical data and guide the extrapolation toward the
				ultra-relativistic limit.
			}
			\label{holo_kc}
		\end{SCfigure}

		The resulting critical momentum is shown in Fig.~\ref{holo_kc}.  
		Panel (a) demonstrates that the holographic convergence radius \(k_c(v)/T\) grows rapidly as \(v\to1\), qualitatively resembling the RTA result before scaled by the Lorentz factor. 
		The decisive difference appears in panel (b): the rescaled quantity \(k_c(v)/(T\gamma(v))\) decreases toward zero.  
		Thus the large growth of \(k_c(v)\) is therefore not the signal; the signal is what remains after the kinematic effect from the Lorentz factor is removed \footnote{The direction dependence of the boosted holographic
			shear-channel convergence radius was studied for moderate boosts,
			$|v|\leq 0.8$, in Ref.~\cite{Hoult:2025htt}. Here we consider the longitudinal $U(1)$ diffusion channel and follow the Lorentz-rescaled convergence radius toward the ultra-relativistic limit.}.

\begin{figure}[t]
			\centering
			\includegraphics[width=.48\textwidth]{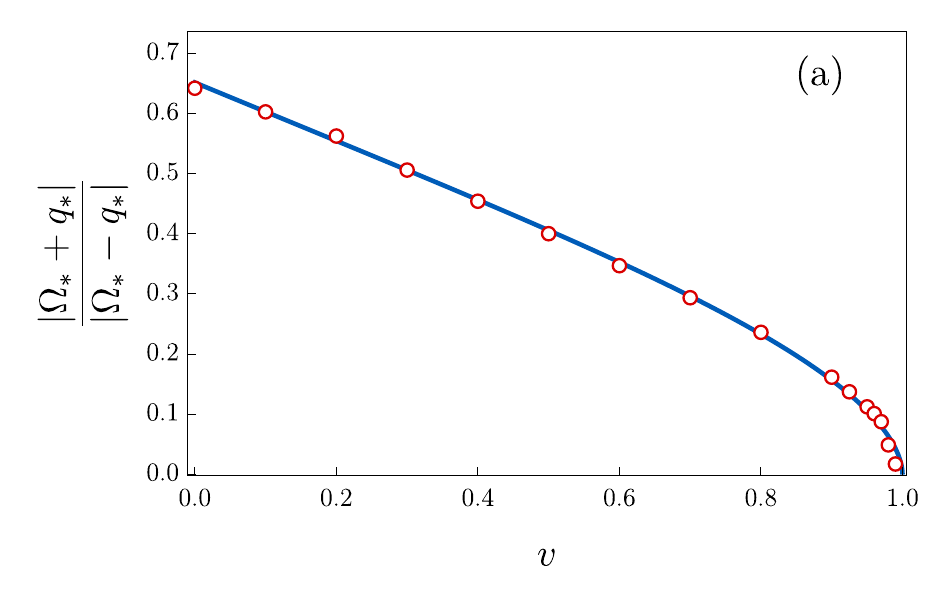}
			\includegraphics[width=.48\textwidth]{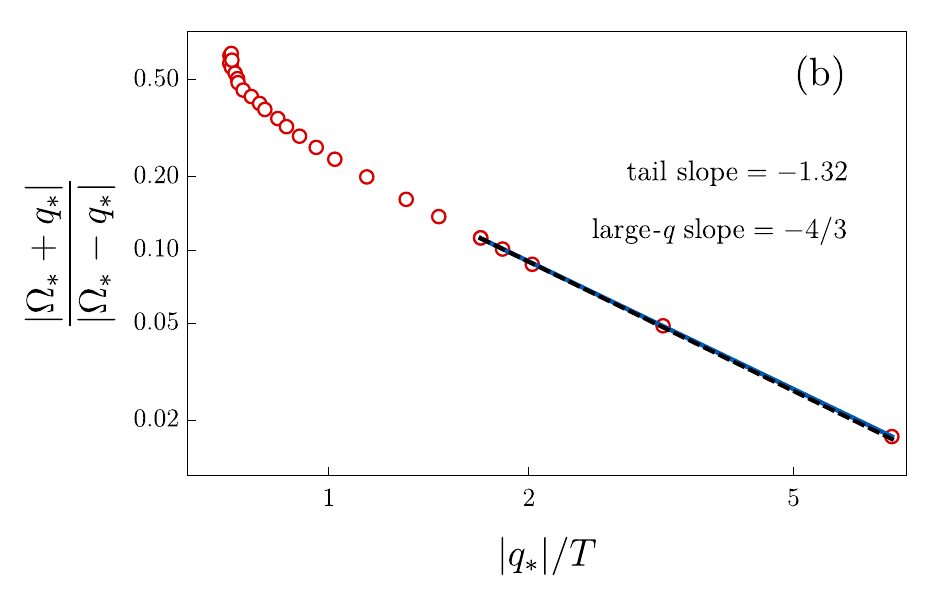}
			\caption{
				RF light-cone diagnostic ratio $\frac{|\Omega_*+q_*|}{|\Omega_*-q_*|}$ for the holographic convergence-limiting branch.
				(a) The ratio $|\Omega_*+q_*|/|\Omega_*-q_*|$ decreases toward zero as $v\to1$, showing that the limiting QNM collision approaches the RF light-cone direction $\Omega_*+q_*=0$.
				(b) The same ratio versus $|q_*|/T$ on a logarithmic scale. 
				The high-$|q_*|$ tail has slope $-1.32$ (blue), close to the large-$q$ asymptotic guide $-4/3$ (dashed).
			}\label{holo_lightcone}
		\end{figure}

		We then map the convergence-limiting point back to the RF variables $(\Omega_*,q_*)$. 
		We find that $\operatorname{Im} q_*(v)\to -\infty$ as $v\to1$. 
		To see which large-momentum direction is being approached, Fig.~\ref{holo_lightcone}(a) shows that $\frac{|\Omega_*+q_*|}{|\Omega_*-q_*|}$ is driven to zero as \(v\to1\).  
		The holographic branch point therefore approaches the RF light-cone direction $\Omega \to -q$.
		
		Figure~\ref{holo_lightcone}(b) gives a quantitative check of this approach; the high-$|q_*|$ tail of the ratio has slope $-1.32$, close to the value $-4/3$ implied by the known large-momentum QNM asymptotics
		\[\Omega_n(q)=\pm q+C_n q^{-1/3}+\ldots\; ,
		\]
		in AdS black-brane backgrounds \cite{Festuccia:2008zx,Fuini:2016qsc}. 
		Since the accessible values of $|q_*|/T$ are moderate, this should be viewed as a consistency check rather than a direct extraction of the asymptotic exponent.

		%______________________________________________________
		\section{Comment on the transverse boost}
		%______________________________________________________
		
		For completeness, let us briefly contrast the longitudinal boost with a boost transverse to the boosted-frame momentum.  
		Taking \(\mathbf v=v\hat x\) and \(\mathbf k=k\hat z\), the RF frequency and momentum components are $\Omega=\gamma\omega$,
		$q_x=-\gamma v\omega=-v\Omega$, and $q_z=k$,
		and hence
		\[
		q^2=k^2+v^2\Omega^2\;,\qquad \omega=\Omega/\gamma \;.
		\]
		Unlike the longitudinal case, there is no relation of the form \(k/\gamma=q+v\Omega\), which is essential to probe the near-light-cone behavior of the hydrodynamic branch in the RF.

		%________________________________________________________
		\section{Discussion and Outlook}
		%________________________________________________________
		We have shown that boosting a thermal state changes the singularity that limits the long-wavelength hydrodynamic expansion. 
		More precisely, we study the analytic continuation of the hydrodynamic dispersion relation in complex momentum, and ask which singularity first obstructs its expansion around $k=0$. 
		At zero boost velocity this obstruction is the familiar nearest nonhydrodynamic structure in the RF spectrum. 
		At finite boost velocity, however, the fixed-$k$ representation of the same microscopic spectrum is reorganized in the sense that the limiting singularity can be selected from a different region of the RF spectral curve, and in the large-boost limit it is controlled by the near-light-cone nonhydrodynamic structure.
		
		Our analysis is restricted to causal and stable relativistic theories, and
		the question we address is the convergence of hydrodynamics rather than the
		onset of instabilities. Boosts and complex momenta also appear in discussions
		of causality, covariant stability, and boosted-frame instabilities
		\cite{Heller:2022ejw,Gavassino:2023mad,Hoult:2023clg}; in particular, they
		have recently been used to study boosted-frame instabilities of
		Gauss--Bonnet black branes outside the allowed causality window
		\cite{Buchel:2026rep}. Related work reconstructed boosted dispersion spectra from rest-frame gradient-expansion data and analyzed additional spurious roots in connection with causality             \cite{Bhattacharyya:2025hjs}.  Here, however, the boost is used in a different way:
		as a probe of how the singularities of a stable microscopic spectrum are
		represented in the fixed-$k$ complex-$\omega$ plane, and hence of which
		singularity controls the boosted hydrodynamic radius of convergence.

		The two examples studied here realize two distinct scenarios of how the light-cone is approached.  
		In resummed RTA diffusion theory, the limiting singularity approaches the RF light-cone with a finite offset given by the relaxation time,  \(\tau k_c(v)/\gamma(v)\to1\). 
		In holographic charge diffusion, the limiting QNM collision is instead driven onto the light-cone itself, \(k_c(v)/(T\gamma(v))\to0\).  
		Thus the boosted convergence radius is not merely a geometric property of the dispersion relation, but a sensitive dynamical probe of how microscopic nonhydrodynamic degrees of freedom obstruct the hydrodynamic branch.

		This turns the convergence problem into a controlled diagnostic tool.  
		At zero boost velocity, the convergence radius probes the obstruction of the hydrodynamic sheet nearest to the origin.  
		By increasing the boost velocity, one continuously changes which region of the RF spectral curve is tested; in the large-boost limit, the test is focused on the near-light-cone regime.

		This should not be interpreted as a universal weak/strong-coupling classification.  
		The leading large-boost behavior reveals that in kinetic theory the light-cone is approached with a finite offset, while in holography that offset vanishes, but other theories may exhibit a finite offset, a vanishing offset, or a different scaling altogether.  
		Subleading large-boost corrections would refine this diagnostic tool by resolving how different microscopic spectral curves approach the light-cone, and by identifying the boosted-frame momentum window in which hydrodynamics remains predictive for perturbations on top of a moving medium.

		\section*{Acknowledgments}
		We thank S.~Grozdanov for valuable discussions. J.K.G.~thanks Daniel Brattan for useful discussions. N.A.~acknowledges support from the National Natural Science Foundation of China under Grant No.~12575142, and from the “Double First-Class” start-up funding of Lanzhou University, China, under Grant No.~561119208.
		D.H.R.~is supported in part by the Deutsche Forschungsgemeinschaft (DFG, German Research Foundation) through the CRC-TR 211 ``Strong-interaction matter under extreme conditions'' –
		project number 315477589–TRR 211.

		\appendix
		\setcounter{equation}{0}
		\appendix
		
		\makeatletter
		\@addtoreset{equation}{section}
		\makeatother
		\renewcommand{\theequation}{\Alph{section}\arabic{equation}}

		%___________________________________________________________________
		\section{Derivation of the boosted RTA spectral equation}
		\label{app_boosted_spectral}
		%___________________________________________________________________
		For completeness, we briefly show how the boosted spectral equation used in the main text follows from the covariant resummed RTA kinetic equation. 
		We consider	a homogeneous background with constant four-velocity
		\begin{equation}			u^\mu=\gamma(1,\mathbf v)\;,
			\qquad
			\gamma=(1-v^2)^{-1/2}\;.
		\end{equation}
		In the absence of external sources, the RTA Boltzmann equation is
		\begin{equation}
			p^\mu\partial_\mu f(x;\mathbf p)
			=
			(p\cdot u)\,
			\frac{f(x;\mathbf p)-f^{(0)}(x;\mathbf p)}{\tau}\;.
			\label{eq:app-boltzmann}
		\end{equation}
		We take a massless classical gas and keep the background temperature fixed.
		With our sign convention, the local equilibrium distribution is
		\begin{equation}
			f^{(0)}(x;\mathbf p)
			=\exp\!\left[\frac{p\cdot u+\mu(x)}{T}\right]\;.
			\label{eq:app-f0}
		\end{equation}
		Since \(u^\mu\) and \(T\) are constant, Eq.~\eqref{eq:app-boltzmann} can be formally solved as
		\begin{equation}
			f=\frac{1}{1-\mathcal D}\,f^{(0)}\;,
			\qquad
			\mathcal D
			=
			\tau\,\frac{p\cdot\partial}{p\cdot u}\;.
			\label{eq:app-formal-sol}
		\end{equation}
		
		The density is fixed by the Landau matching condition. 
		With the present sign convention it is convenient to write
		\begin{equation}
			n
			\equiv
			-\int\frac{d^3p}{(2\pi)^3}\frac{p\cdot u}{p^0}\,
			f(x;\mathbf p)
			=
			-\int\frac{d^3p}{(2\pi)^3}\frac{p\cdot u}{p^0}\,
			f^{(0)}(x;\mathbf p)\;.
			\label{eq:app-landau}
		\end{equation}
		Substituting Eq.~\eqref{eq:app-formal-sol} into this condition gives
		\begin{equation}
			\boxed{
				\int\frac{d^3p}{(2\pi)^3}
				\frac{p\cdot u}{p^0}\,
				\frac{\mathcal D}{1-\mathcal D}
				f^{(0)}(x;\mathbf p)=0
			} \;.
			\label{eq:app-master-before-boost}
		\end{equation}
		The overall sign is immaterial in this equation.
		
		In the local rest frame, \(f^{(0)}\) is isotropic in momentum space, and the		phase-space integral reduces to the angular average used in Ref.~\cite{Abbasi:2024pwz}. 
		In the boosted frame, however, \(f^{(0)}\) is not isotropic in momentum \(\mathbf p\). 
		It is therefore useful to perform a boost to variables \(\tilde{\mathbf p}\) defined by
		\begin{equation}
			\begin{split}
				-p\cdot u
				=
				\gamma(p^0-\mathbf p\cdot\mathbf v)
				&\equiv \tilde p^0\;,
				\\
				\mathbf p
				+(\gamma-1)(\mathbf p\cdot\hat{\mathbf v})\hat{\mathbf v}
				-\gamma p^0\mathbf v
				&\equiv \tilde{\mathbf p}\; .
			\end{split}
			\label{eq:app-momentum-boost}
		\end{equation}
		The Lorentz-invariant measure is unchanged,
		\begin{equation}
			\frac{d^3p}{p^0}
			=
			\frac{d^3\tilde p}{\tilde p^0}\;,
		\end{equation}
		and the local-equilibrium distribution becomes isotropic,
		\begin{equation}
			\label{eq:S9}
			f^{(0)}(x;\mathbf p)
			=
			\exp\!\left[\frac{p\cdot u+\mu(x)}{T}\right]
			\longrightarrow
			f^{(0)}(x;\tilde{\mathbf p})
			=
			\exp\!\left[-\frac{\tilde p^0-\mu(x)}{T}\right]\;.
		\end{equation}
		The differential operator transforms covariantly. 
		Dropping the tilde after the change of variables, one obtains
		\begin{equation}\label{D_box}
			\boxed{
				\mathcal D
				=
				-\tau\left\{
				\gamma(1+\hat{\mathbf p}\cdot\mathbf v)\,\partial_t
				+
				\left[
				\hat{\mathbf p}
				+(\gamma-1)(\hat{\mathbf p}\cdot\hat{\mathbf v})\hat{\mathbf v}
				+\gamma\mathbf v
				\right]\cdot\nabla
				\right\}
			} \;,
		\end{equation}
		with $\hat{\mathbf{p}} \equiv \mathbf{p}/p^0$ and  
		\begin{equation}\label{f_0}
			f^{(0)}(x;\mathbf p)
			=
			\exp\!\left[-\frac{p^0-\mu(x)}{T}\right]\;.
		\end{equation}
		
		In these variables the phase-space integral factorizes into a radial integral and an angular average. 
		Defining 
		$\int_{\hat{\mathbf p}}		\equiv \frac{1}{4\pi}\int d\Omega_{\hat{\mathbf p}},$ and using the definition of \(n(x)\), Eq.~\eqref{eq:app-master-before-boost} reduces to
		\begin{equation}
			\boxed{
				\int_{\hat{\mathbf p}}
				\frac{\mathcal D}{1-\mathcal D}\,n(x)=0
			} \;.
			\label{eq:app-final-position}
		\end{equation}
		This is the boosted analogue of the resummed RTA diffusion equation used in the rest frame.
		
		Passing to Fourier space,
		$n(x) =	\int d\omega\,d^3k\,
		e^{-i\omega t+i\mathbf k\cdot\mathbf x}
		n(\omega,\mathbf k),$
		we find
		\begin{equation}
			\int_{\hat{\mathbf p}}
			\frac{\bar{\mathcal D}}{1-\bar{\mathcal D}}\,
			n(\omega,\mathbf k)=0\;,
			\label{eq:app-spectral-pre}
		\end{equation}
		where
		\begin{equation}
			\bar{\mathcal D}
			=
			i\tau\left\{
			\gamma(\omega-\mathbf v\cdot\mathbf k)
			+
			\hat{\mathbf p}\cdot
			\left[
			\gamma\omega\mathbf v
			-\mathbf k
			-(\gamma-1)\hat{\mathbf v}
			(\hat{\mathbf v}\cdot\mathbf k)
			\right]
			\right\}\;.
			\label{eq:app-Dbar}
		\end{equation}
		Equation~\eqref{eq:app-spectral-pre}, together with		Eq.~\eqref{eq:app-Dbar}, is the boosted spectral equation of the resummed RTA diffusion theory used	in the main text.

		%________________________________________________________
		\section{Loss of invertibility and the double-root condition}
		\label{app_loss_of_invertl}
		%________________________________________________________
		
		In this section we show that, in the longitudinal channel, the loss-of-invertibility condition for the boosted hydrodynamic branch,
		\[
		\frac{dk}{dq}=0\; ,
		\]
		is equivalent, in the generic case, to the usual double-root characterization of a branch point of the spectral equation,
		\[
		F(\omega,k;v)=0\;,
		\qquad
		\partial_\omega F(\omega,k;v)=0 \;.
		\]
		The statement is local and assumes that we work on a chosen analytic sheet of the spectral function.
		
		Along the boosted hydrodynamic branch, parametrized by Eqs.~\eqref{boosted_parametric_branch}, the exact dispersion relation satisfies
		\begin{equation}
			F\bigl(\omega(q;v),k(q;v);v\bigr)=0 \;.
			\label{app-F-on-branch}
		\end{equation}
		Differentiating with respect to \(q\) gives
		\begin{equation}
			0=\frac{dF}{dq}
			=
			\partial_\omega F\,\frac{d\omega}{dq}
			+
			\partial_k F\,\frac{dk}{dq}\;.
			\label{app-dFdq}
		\end{equation}
		For \(\mathbf k\parallel \mathbf v\), Eqs.~\eqref{boosted_parametric_branch} imply
		\begin{equation}
			\frac{dk}{dq}
			=
			\gamma\bigl[1+v\,\Omega'(q)\bigr]\;,
			\qquad
			\frac{d\omega}{dq}
			=
			\gamma\bigl[v+\Omega'(q)\bigr]\;,
			\label{app-dk-dw-dq}
		\end{equation}
		where \(\Omega(q)\) denotes the corresponding RF branch.
		
		At the point $q_*$ where the map \(q\mapsto k(q;v)\) ceases to be invertible, one has
		\begin{equation}
			\frac{dk}{dq}(q_*;v)=0
			\qquad\Longleftrightarrow\qquad
			1+v\,\Omega'(q_*)=0 \;.
			\label{app-jac-zero}
		\end{equation}
		Therefore
		\begin{equation}
			\Omega'(q_*)=-\frac{1}{v}\;,
			\qquad
			\left.\frac{d\omega}{dq}\right|_{q_*}
			=
			\gamma\left(v-\frac{1}{v}\right)
			=
			-\frac{1}{\gamma v}\neq 0
			\qquad (0<v<1)\;.
			\label{eq:app-dwdq-nonzero}
		\end{equation}
		Using Eq.~\eqref{app-dFdq}, we then find
		\begin{equation}
			\frac{dk}{dq}(q_*;v)=0
			\qquad\Longrightarrow\qquad
			\partial_\omega F\bigl(\omega_*,k_*;v\bigr)=0 \; ,
			\label{eq:app-jac-implies-droot}
		\end{equation}
		with \((\omega_*,k_*)=(\omega(q_*;v),k(q_*;v))\). 
		Since the point lies on the branch, \(F(\omega_*,k_*;v)=0\) also holds.
		
		Conversely, suppose that on the same branch
		\begin{equation}
			F(\omega_*,k_*;v)=0\;,
			\qquad
			\partial_\omega F(\omega_*,k_*;v)=0\;,
			\label{eq:app-double-root}
		\end{equation}
		and furthermore
		\begin{equation}
			\partial_k F(\omega_*,k_*;v)\neq 0 \;.
			\label{eq:app-generic-pk}
		\end{equation}
		This is the generic square-root branch-point case. 
		Equation~\eqref{app-dFdq} then implies
		\begin{equation}
			\frac{dk}{dq}(q_*;v)=0 .
			\label{eq:app-droot-implies-jac}
		\end{equation}
		Therefore, for \(0<v<1\) and in the generic case
		\(\partial_k F\neq0\),
		\begin{equation}
			\boxed{
				\frac{dk}{dq}(q_*;v)=0
				\quad\Longleftrightarrow\quad
				F(\omega_*,k_*;v)=0\;,
				\qquad
				\partial_\omega F(\omega_*,k_*;v)=0 \;.
			}
			\label{eq:app-equivalence}
		\end{equation}
		
		The RF case \(v=0\) is exceptional. 
		In that case \(k=q\), so \(dk/dq=1\), and the nearest obstruction is not produced by the boosted projection. 
		It instead comes directly from the singularities of the RF  hydrodynamic branch \(\Omega(q)\).
		
		%___________________________________________________________
		\section{Real-momentum QNM spectra under a boost}
		%___________________________________________________________
		In this section we show the spectra at fixed real boosted-frame momentum. These plots are not used to determine the convergence radius; the latter requires analytic continuation to complex \(k\), as discussed in the main text. 
		Their purpose is instead to give a direct view of how the hydrodynamic mode, the kinetic-theory branch cut, and the holographic QNM tower appear in boosted-frame frequency variables.
		
		%___________________________________________________________
		\subsection{Resummed RTA diffusion theory}
\begin{figure}[t]
			\centering				\includegraphics[width=.5\textwidth]{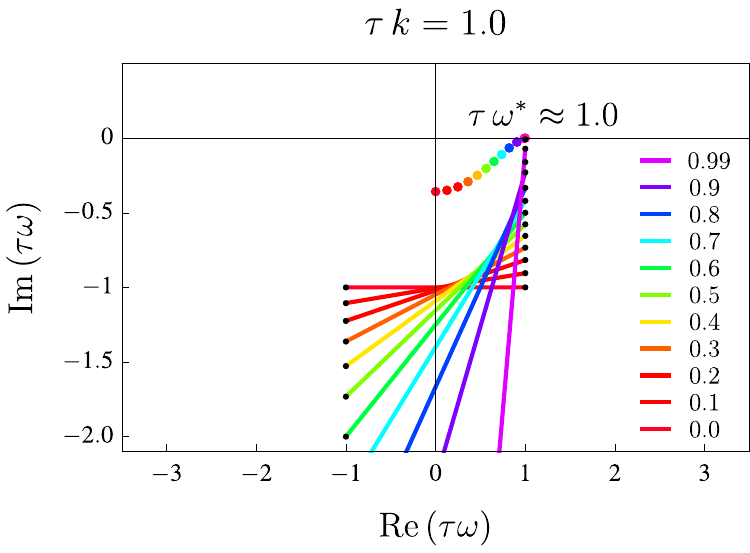}
			\vspace{-2.ex}
			\caption{
				Physical-sheet analytic structure of the longitudinal spectrum at
				fixed $\tau k=1$. Colored dots denote the hydrodynamic mode, while colored
				segments denote the corresponding nonhydrodynamic logarithmic branch cuts in
				the boosted complex-$\omega$ plane; black dots mark the logarithmic branch
				points. The same color corresponds to the same boost velocity. The analytic
				continuation is performed on the physical branch, with
				$Q(\omega)=\gamma(k-v\omega)$ for $k>0$. As the boost is increased, the
				hydrodynamic mode is displaced toward the convective line $\omega\simeq vk$.
				At the same time, the logarithmic cut is deformed asymmetrically: one endpoint
				approaches the light-cone point $\omega=k$ from below, while the other moves
				deeper into the lower half-plane.
			}
			\label{RTA_spectral}
		\end{figure}

		%___________________________________________________________
		Figure~\ref{RTA_spectral} shows the longitudinal resummed RTA diffusion spectrum at fixed $\tau k=1$. 
		The colored dots denote the hydrodynamic  mode on the physical sheet, while the  straight-lined colored segments denote the nonhydrodynamic branch cuts. 
		The black endpoints are the logarithmic branch points of the angular integral.
		
		The origin of the cut is most transparent in the RF. 
		It is produced by the continuum of noncollective particle excitations which differ by their velocity direction. 
		Equivalently, it comes from the angular integral over
		\begin{equation}
			\mu=\hat{\mathbf p}\cdot\hat{\mathbf q}\;,
			\qquad
			-1\leq \mu\leq 1\; .
		\end{equation}
		This is the same physical mechanism emphasized in the free-streaming picture of Kurkela and Wiedemann~\cite{Kurkela:2017xis}.
		In the latter reference, the cut comes from the continuum of particle propagation directions, and the branch points are associated with the endpoints of this angular continuum.

		In the boosted fixed-$k$ spectral problem, however, the branch points in the complex $\omega$-plane are not obtained by simply Lorentz-transforming the RF endpoints. 
		The reason is that the RF endpoints
		\[
		\tau\Omega_{\rm bp}^{\pm}=\pm \tau q-i
		\]
		are points on the endpoint curves of the RF angular integral, parametrized by the RF momentum $q$. 
		Lorentz-transforming those points would give the boosted image of those RF endpoint curves. 
		This is not the operation used in Fig.~\ref{RTA_spectral}. 
		There, the boosted momentum $k$ is fixed from the outset, and the spectrum is viewed as a function of the remaining complex variable $\omega$.
		
		Thus the correct procedure is to start from the boosted angular integral itself.
		We first express the RF variables entering the logarithm in terms of the boosted	variables $(\omega,k)$, then keep $k$ fixed, and finally locate the zeros of the argument of the resulting logarithm. 
		These endpoint zeros are the logarithmic branch points of the boosted fixed-$k$ Green's function. 
		They are therefore not generated by the turning-point condition used in the main text for the boosted convergence scale; they are endpoint singularities of the logarithm itself.
		
		For the longitudinal configuration considered here, the physical-sheet continuation is obtained by writing, for $k>0$,
		\begin{equation}
			Q(\omega)=\gamma(k-v\omega)\;,\qquad
			\Omega(\omega)=\gamma(\omega-vk)\;,
		\end{equation}
		rather than by continuing the real-axis magnitude $|k-v\omega|$. The two
		logarithmic endpoints are then determined by
		\begin{equation}
			{i\over \tau}-Q(\omega)+\Omega(\omega)=0\;,
			\qquad
			{i\over \tau}+Q(\omega)+\Omega(\omega)=0 \;.
		\end{equation}
		Solving these equations gives
		\begin{equation}
			\tau\omega_{\rm bp}^{\pm}
			=
			\pm \tau k
			-
			{i\over \gamma(1\pm v)} \;.
			\label{eq:RTA_boosted_branch_points}
		\end{equation}

		Thus the two endpoints are affected by different Doppler factors. 
		As $v\to1$, the endpoint at positive real frequency approaches the light-cone
		point,
		\begin{equation}
			\omega_{\rm bp}^{+}\to k-i0^{+}\;,
		\end{equation}
		whereas the endpoint at negative real frequency moves deeper into the lower
		half-plane,
		\begin{equation}
			\omega_{\rm bp}^{-}\to -k-i\infty\; .
		\end{equation}
		This provides a simple illustration of the general point emphasized in the
		main text.

		%%%%%%%%%%%%%%%%%%%%%%%%%%%%%%%%%%
		%%%%%%%%%%%%%%%%%%%%%%%%%%%%%%%%%%

		The hydrodynamic mode behaves differently. 
		It is a collective zero of the full spectral equation and is not constrained by the endpoint symmetry of the logarithmic cut. 
		For a longitudinal boost, we have 
		\begin{equation}
			\omega=vk+\frac{\Omega}{\gamma}\;.
		\end{equation}
		Therefore a collective excitation whose RF frequency remains finite is compressed, in boosted variables, into a band of width $O(1/\gamma)$ around the convective line
		\begin{equation}
			\omega\simeq vk\; .
		\end{equation}
		This compression effect is physical: it is a consequence of the fact that causality requires that the poles of the retarded correlator are located in the lower half of the complex frequency plane. 
		In a causal and stable theory, such as ours, a boost cannot lead to the poles crossing into the upper half plane, so they must accumulate near the convective line.
		In particular,
		\begin{equation}
			\mathrm{Im}\,\omega=\frac{\mathrm{Im}\,\Omega}{\gamma}\;.
		\end{equation}
		This explains the near-real-axis appearance of the hydrodynamic mode at large
		boost velocity. 

		\subsection{Holography}
\begin{figure*}[t]
			\centering
			\vspace{-1.0ex}
			\begin{minipage}[t]{0.45\textwidth}
				\centering
				\begin{overpic}[width=\linewidth]{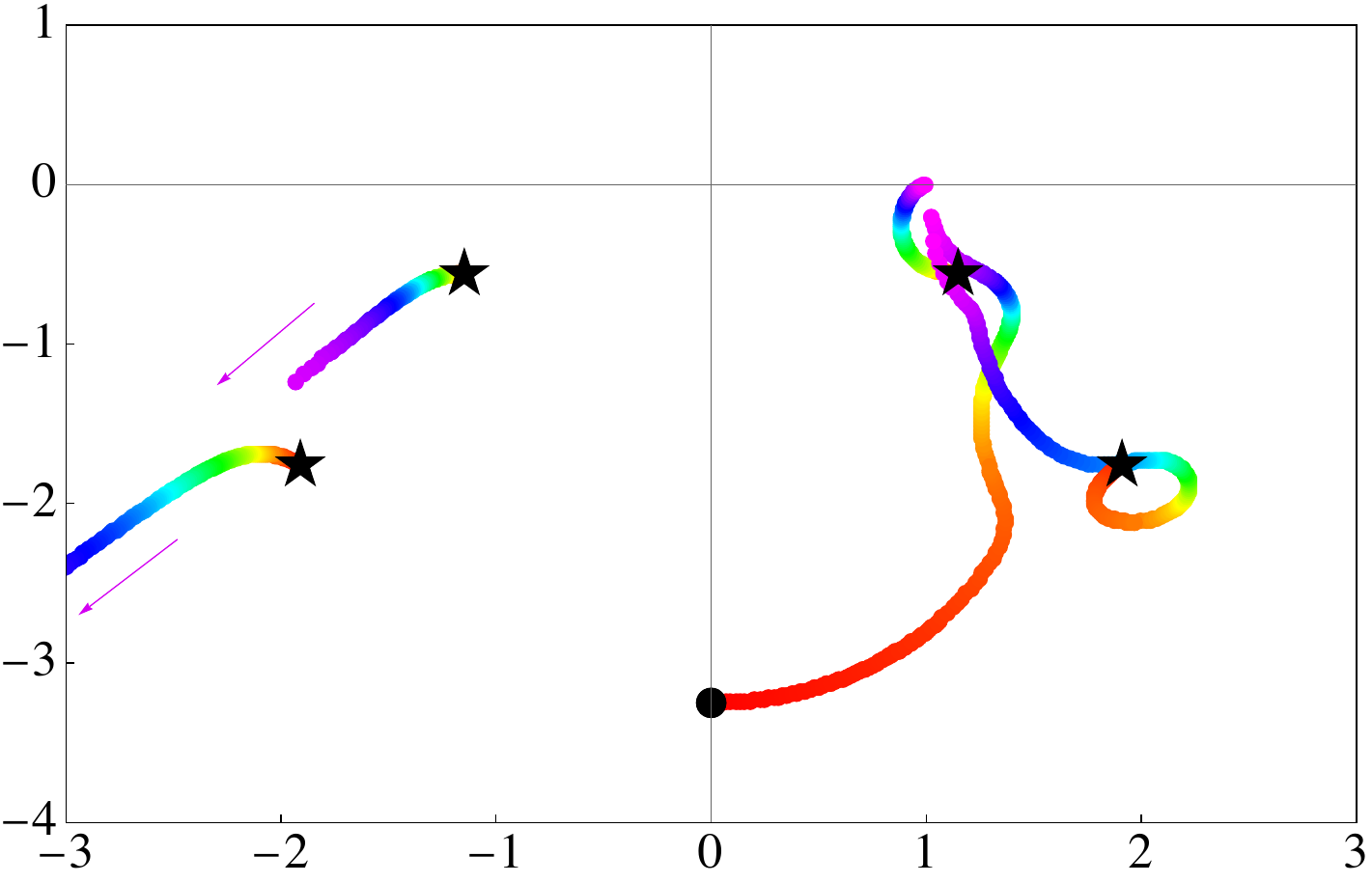}
					\put(47,-5){ \large $\mathrm{Re}\,\frac{\omega}{T}$}
					\put(-10,30){ \rotatebox{90}{\large $\mathrm{Im}\,\frac{\omega}{T}$}}
					\put(75,55){{\large $\frac{k}{T}=1$}}
				\end{overpic}
			\end{minipage}
			\vspace{1.ex}
			\caption{
				Holographic QNM trajectories at fixed real boosted-frame momentum $k/T=1$ as
				the boost velocity is increased. The black markers show the corresponding RF
				spectrum: the black dot denotes the diffusive pole, while the stars denote
				nonhydrodynamic QNMs. As the boost is turned on, each mode traces a trajectory
				in the complex $\omega$-plane; the color gradient indicates increasing $v$.
				The apparent pile-up of trajectories near the convective line
				$\omega\simeq vk$ is a physical-sheet compression effect. Modes with finite RF
				frequency $\Omega=O(T)$ obey $\omega-vk=\Omega/\gamma$ and are therefore
				squeezed into a band of width $\Delta\omega\sim T/\gamma$ in boosted variables.
				The hydrodynamic trajectory is identified by continuity from the diffusive pole
				at small RF momentum.	}
			\label{fig:holo_spec_Epar_k1}
		\end{figure*}

		%	The holographic QNM
		%___________________________________________________________
		Figure~\ref{fig:holo_spec_Epar_k1} shows the holographic QNM trajectories at fixed real boosted-frame momentum, \(k/T=1\), as the boost velocity is increased.
		The most prominent feature is the apparent pile-up of several trajectories near the convective line
		\[
		\omega\simeq vk \;.
		\]
		This accumulation is similar to the compression effect in RTA diffusion theory discussed above.
		It is not related to the convergence-limiting singularity discussed in the main text.
		The convergence radius is determined only after analytic continuation to complex $k$. 
		Nevertheless, the plot is useful because it shows how a large portion of the RF QNM spectrum is compressed when represented in boosted-frame frequency variables.
		
		For a longitudinal boost, the RF and boosted variables are related by
		\begin{equation}
			\Omega=\gamma(\omega-vk)\;,
			\qquad
			q=\gamma(k-v\omega)\;,
			\label{eq:holo_boost_forward}
		\end{equation}
		or, equivalently,
		\begin{equation}
			\omega=vk+\frac{\Omega}{\gamma}.
			\label{eq:holo_boost_inverse}
		\end{equation}
		Thus any QNM whose RF frequency remains finite, $\Omega=O(T)$, is mapped in the boosted frame to
		\begin{equation}
			\omega-vk=O\!\left(\frac{T}{\gamma}\right)\;,
			\qquad
			\mathrm{Im}\,\omega=O\!\left(\frac{T}{\gamma}\right)\;.
			\label{eq:holo_pileup_scaling}
		\end{equation}
		As $v\to1$, such modes are squeezed into a narrow band of width $\Delta\omega\sim T/\gamma$ around the convective line. 
		This is the origin of the pile-up seen on the right-hand side of Fig.~\ref{fig:holo_spec_Epar_k1}.
		It is simply the lab-frame time dilation of modes that have finite decay rates in the plasma rest frame.

		%%%%%%%%%%%%%%%%%%%%%%%%%%%%%%%%%%
		%%%%%%%%%%%%%%%%%%%%%%%%%%%%%%%%%%

		The same observation also explains why the hydrodynamic trajectory can appear to join the pile-up. 
		The label ``hydrodynamic'' is tied to the small-$q$ RF	regime, where the mode is long-lived and is well approximated by its hydrodynamic expansion. 
		In Fig.~\ref{fig:holo_spec_Epar_k1}, however, we keep the boosted momentum $k$ fixed while increasing $v$. 
		The corresponding RF momentum $q$, given by Eq.~\eqref{eq:holo_boost_forward}, is not held fixed and need not remain small. 
		Once the trajectory is pushed	to moderate or large RF momentum, the hydrodynamic mode should be understood as	the analytic continuation of the lowest QNM rather than as a mode described by the small-$q$ hydrodynamic expansion. 
		If its RF frequency remains finite along the part of the trajectory being sampled, Eq.~\eqref{eq:holo_boost_inverse} compresses it toward $\omega\simeq vk$ in the same way as the nonhydrodynamic 	modes.
		
		The trajectories that remain on the left-hand side of the plot have a different RF interpretation. 
		At fixed positive boosted momentum, a point with	$\mathrm{Re}\,\omega<0$ gives
		\[
		\mathrm{Re}\,\Omega
		=
		\gamma(\mathrm{Re}\,\omega-vk)
		<
		-\gamma vk \;,
		\]
		which is parametrically large and negative as $v\to1$. 
		Such points therefore sample a different large-frequency region of the RF spectrum, rather than the	finite-$\Omega$ sector compressed near the convective line. 
		The right-hand pile-up is thus not a universal attraction of all trajectories to 	$\omega=vk$; it is the boosted image of the part of the RF spectrum whose frequency remains $O(T)$.

		%__________________________________________________________
		\section{Large-boost limit of the RTA critical point}
		\label{large_limit}
		%___________________________________________________________
		In the main text we used the fact that, in the large-boost limit, the RTA critical point selected by the pole-connected branch is driven to large negative imaginary momentum. 
		Here we justify this statement directly
		from the critical-point equation. 
		The condition is $\Omega'(q_\ast)=-1/v$; hence, as $v\to1$, any finite lower-half-plane limit would have to satisfy $\Omega'(q)+1=0$. 
		We show below that no such finite solution exists.
		
		Let \(z=\tau q\) and \(s=e^{-2iz}\). For the RTA branch,
		\begin{equation}
			\Omega'(q)+1
			=
			{s(4iz-2+2s)\over(1-s)^2}\;.
		\end{equation}
		In the large-boost limit the critical-point condition requires $\Omega'(q_\ast)+1\to0$. 
		A finite lower-half-plane limit, namely $s\ne 0$, would therefore require, away from the poles,
		\begin{equation}
			e^{-2iz}=1-2iz\; .
		\end{equation}
		Defining \(w=-2iz\), this becomes
		\begin{equation}
			e^w=1+w \;.
		\end{equation}
		For ${\rm Im}\,z\le0$, one has ${\rm Re}\,w\le0$. 
		Writing	$w=a+ib$ with $a\le0$, the imaginary part gives
		\begin{equation}
			b=e^a\sin b\; .
		\end{equation}
		Since $e^a\le1$, this equation has only the solution $b=0$. 
		The real part then gives
		\begin{equation}
			e^a=1+a\; ,
		\end{equation}
		whose only solution is $a=0$. 
		Thus the only solution with ${\rm Re}\,w\le0$ is $w=0$, i.e., $z=0$. 
		However $z=0$ is a removable point of the above expression and directly $\Omega'(0)+1=1$. 
		Hence there is no finite lower-half-plane solution of $\Omega'(q)+1=0$. 
		The relevant critical point is therefore driven to
		\begin{equation}
			e^{-2iz_\ast}\to0\;,\qquad {\rm Im}\,z_\ast\to-\infty \;.
		\end{equation}

		%________________________________________________________________
		\section{Independent Taylor-series check of the convergence radius in the longitudinal channel}
		\label{app_taylor-check}
		%________________________________________________________________
		As an independent check of the branch-point construction used for the RTA analysis in the main text, we also extracted the convergence radius directly from the large-order behavior of the hydrodynamic Taylor series around $k=0$ at fixed boost velocity $v$.
		We use the same notation as in the main text. 
		In the longitudinal channel, $\mathbf{k}\parallel\mathbf{v}$, the exact RF hydrodynamic mode is parametrized by Eq.~(\ref{branch}). 
		Expanding the hydrodynamic branch around $k=0$, we write
		%%%%%%%%%%%%%%%%%%
		\begin{equation}
			\omega(k;v)=\sum_{n\ge 1} a_n(v)\,k^n \;.
			\label{eq:B1}
		\end{equation}
		%%%%%%%%%%%%%%%%%%
		The coefficients $a_n(v)$ were generated to high order by expanding the parametric map~\eqref{boosted_parametric_branch} around $q=0$, inverting the resulting series for $k(q;v)$, and substituting the inverse into $\omega(q;v)$.
		
		As pointed out in the main text, for \(v\neq0\) the convergence-limiting	singularity is generically a square-root branch point. 
		%	Let \(q_*\) denote a
		%	solution of the loss-of-invertibility condition,
		%	\[
		%	\frac{dk}{dq}(q_*;v)=0 .
		%	\]
		%	For a generic branch point one has \(k''(q_*;v)\neq0\), while
		%	\(\omega'(q_*;v)\neq0\). Expanding the parametric map near \(q=q_*\) gives
		%	\[
		%	k-k_*=
		%	\frac{1}{2}k''(q_*;v)(q-q_*)^2+\cdots ,
		%	\qquad
		%	\omega-\omega_*=
		%	\omega'(q_*;v)(q-q_*)+\cdots .
		%	\]
		%	Eliminating \(q-q_*\), one obtains
		%	\[
		%	\omega(k)-\omega_*
		%	=
		%	B_*
		%	\sqrt{k-k_*}
		%	+\cdots ,
		%	\]
		%	with a nonzero constant \(B_*\). Thus the analytically continued
		%	hydrodynamic dispersion relation has a square-root branch point at \(k=k_*\).
		%	This is the singularity that controls the large-order behavior of the Taylor
		%	coefficients around \(k=0\).
		%	
		To connect this local square-root singularity to the large-order Taylor coefficients, we use the standard singularity-analysis (Darboux's method) relation between	nearest singularities and coefficient asymptotics~\cite{Grozdanov:2022npo,Flajolet:1990,Hunter:1980}. 
		Let the singular part near $k=k_*$ be written, up to terms analytic at $k_*$, as
		\begin{equation}
			\omega_{\rm sing}(k)
			=
			B_*
			\left(1-\frac{k}{k_*}\right)^{1/2}\;.
		\end{equation}
		The relevant coefficient asymptotics follows from
		\begin{equation}
			[k^n]\left(1-\frac{k}{k_*}\right)^\alpha
			\sim
			\frac{k_*^{-n}}{\Gamma(-\alpha)}\,n^{-\alpha-1}\;,
			\qquad n\to\infty \;,
		\end{equation}
		valid for noninteger $\alpha$ when this singularity is among the nearest ones to the origin. 
		Here, by $[k^n]\cdots$, we mean the coefficient of $k^n$ in the Taylor expansion of $\cdots$ about $k=0$. 
		For the square-root case $\alpha=1/2$, this gives
		\begin{equation}
			[k^n]\left(1-\frac{k}{k_*}\right)^{1/2}
			\sim
			\frac{k_*^{-n}}{\Gamma(-1/2)}\,n^{-3/2}\;.
		\end{equation}
		Therefore a single square-root branch point contributes
		\begin{equation}
			a_n^{(*)}(v)
			\sim
			\mathcal{A}_*\, k_*^{-n}\, n^{-3/2}\;,
			\qquad \mathcal A_* \equiv \frac{\mathcal B_*}{\Gamma(-1/2)}\; .
		\end{equation}
		In the cases relevant below, the nearest singularities are either a single dominant square-root branch point or a pair of singularities with the same modulus. 
		The latter situation is visible in the high-boost data through the oscillatory modulation of the finite-order coefficients.
		The relevant pair of singularities is fixed by the real-valuedness  condition of the retarded correlator. 
		The latter is the statement that, if $F(\omega,k;v)=0$ then $F(-\omega^*,-k^*;v)=0 .$
		Thus, if one limiting singularity of the hydrodynamic mode is located at
		\[
		k_1=k_c e^{i\theta}\;,
		\]
		the symmetry-related partner is
		\[
		k_2=-k_1^*=k_c e^{i(\pi-\theta)}\; .
		\]
		Moreover, this symmetry relates the local square-root expansions. 
		If near	\(k_1\)
		\[
		\omega(k)=\omega_1+B_1\left(1-\frac{k}{k_1}\right)^{1/2}+\ldots \;,
		\]
		then near \(k_2\)
		\[
		\omega(k)=-\omega_1^*
		-B_1^*\left(1-\frac{k}{k_2}\right)^{1/2}+\ldots\; ,
		\]
		up to the conventional choice of square-root branch.				%%%%%%%%%%%%%%%%%%%%%%%%
		\begin{figure}[t]
			\centering
			\includegraphics[width=.48\textwidth]{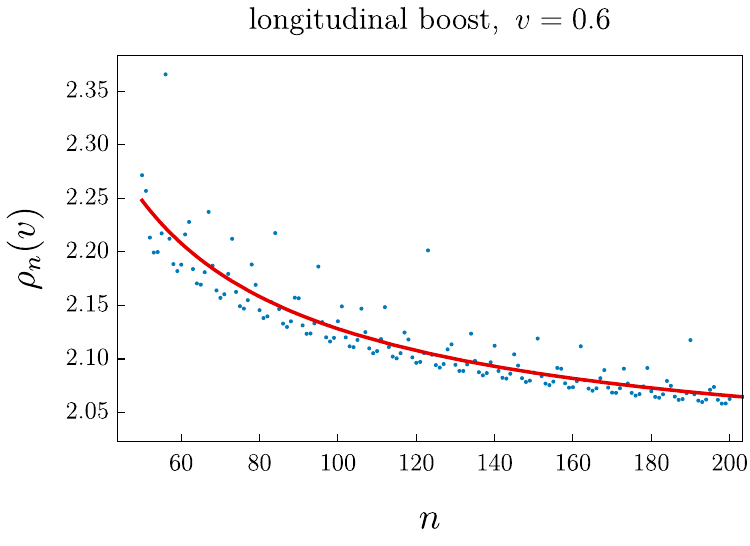}\,		\includegraphics[width=.48\textwidth]{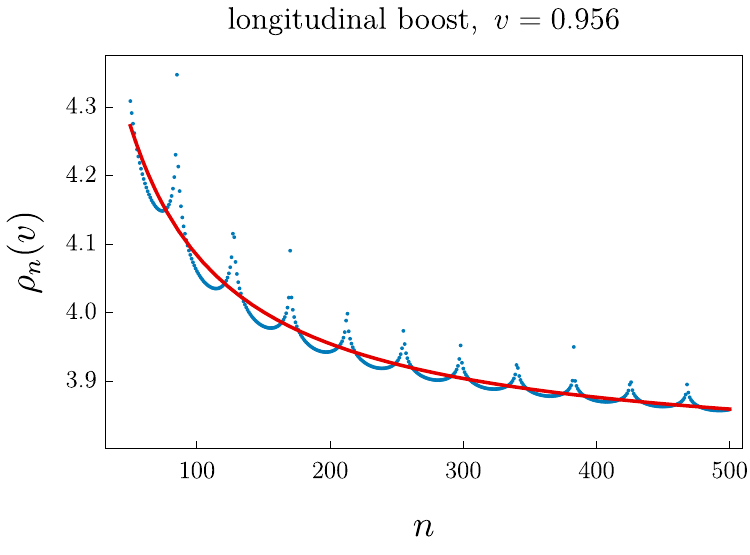}
			\caption{
				Finite-order root estimator
				\(
				\rho_n(v)=|a_n(v)|^{-1/n}
				\)
				constructed from the Taylor coefficients of the longitudinal hydrodynamic dispersion relation, shown for \(v=0.6\) and \(v=0.956\). Blue points are the finite-order data. The red curves show the smooth large-\(n\) envelope fit based on Eq.~\eqref{eq:root_estimator_fit}.  For a moderate boost the approach to the asymptotic value \(k_c(v)\) is nearly monotonic, whereas at larger boost a clear oscillatory modulation develops. 
				The extrapolated large-$n$ limit agrees with the convergence radius obtained independently from the singularity structure in the complex momentum plane. A quantitative comparison, including the two velocities shown here and one additional representative boost, is given in Table~\ref{taylor_check}.}
			\label{comparison}
		\end{figure}
		%%%%%%%%%%%%%%%%%%%%%%%%
		Since \(\Gamma(-1/2)\) is real, the corresponding large-order amplitudes satisfy \begin{equation}
			\mathcal A_2=-\mathcal A_1^* \;.
		\end{equation} 
		The leading contribution of the pair to the Taylor coefficients is therefore
		\begin{equation}
			a_n(v) \sim n^{-3/2} \left[ \mathcal A_1 k_1^{-n} - \mathcal A_1^* k_2^{-n} \right]\;, \qquad n\to\infty\; .
		\end{equation} 
		Using \(k_1=k_c e^{i\theta}\) and \(k_2=k_c e^{i(\pi-\theta)}\), this can be written as
		\begin{equation} \label{a_n}
			a_n(v) \sim k_c^{-n} n^{-3/2} \left[ \mathcal A_1 e^{-in\theta} - (-1)^n \mathcal A_1^* e^{in\theta} \right]\;,
		\end{equation}
		
		Equivalently, writing \(A_1=|A_1|e^{i\alpha}\), the parity factor in
		Eq.~\eqref{a_n} gives
		\begin{equation}
			a_n(v)\sim
			\begin{cases}
				2i|A_1|k_c^{-n}n^{-3/2}\sin(\alpha-n\theta)\;,
				& n\ {\rm even}\;,\\[3pt]
				2|A_1|k_c^{-n}n^{-3/2}\cos(\alpha-n\theta)\;,
				& n\ {\rm odd}\;.
			\end{cases}
			\label{eq:oscillatory_coeff_asymptotics}
		\end{equation}
		The symmetry-related pair produces an oscillatory modulation of the
		finite-order estimator, including possible near-cancellations.
		
		The oscillatory prefactor in Eq.~\eqref{eq:oscillatory_coeff_asymptotics}
		makes simple ratio tests unreliable;	successive coefficients can have strong phase/sign fluctuations and occasional	near-cancellations. 
		We therefore use the finite-order root estimator
		\begin{equation}
			\rho_n(v)\equiv |a_n(v)|^{-1/n}\;.
		\end{equation}
		By the Cauchy--Hadamard formula \cite{AhlforsComplexAnalysis},
		\begin{equation}
			\frac{1}{k_c(v)}
			=
			\limsup_{n\rightarrow\infty}|a_n(v)|^{1/n},
		\end{equation}
		or equivalently,
		\begin{equation}
			k_c(v)
			=
			\liminf_{n\rightarrow\infty}\rho_n(v).
		\end{equation}
		Thus, the sequence $\rho_n(v)$ provides natural finite-order root
		estimates of the radius of convergence.

		If the dominant large-order behavior is
		\begin{equation}
			a_n(v)\sim {\cal A}\,k_c^{-n}n^{-3/2}
		\end{equation}
		then the slowly varying trend of the root estimator behaves as
		\begin{equation}
			\rho_n(v)
			=
			k_c(v)
			\exp\left[
			\frac{\frac32\log n-\log|{\cal A}|+o(1)}{n}
			\right]\;.
			\label{eq:root-asymptotics}
		\end{equation}
		This motivates the fitting form
		\begin{equation}
			\rho_n(v)
			=
			k_c(v)
			\left(
			1+\frac{\frac32\log n+b}{n}
			+\frac{c}{n^2}
			\right)\;,
			\label{eq:root_estimator_fit}
		\end{equation}
		where \(b\) absorbs the nonuniversal amplitude and \(c\) parametrizes	subleading corrections. 
		The coefficient of \(\log n/n\) is fixed to \(3/2\), as dictated by the square-root exponent.

		Applying this procedure at representative boost velocities $v=0.6$, $0.8$, and $0.956$, we obtained values of $k_c(v)$ consistent with those extracted in the main text from the branch-point/Jacobian construction. 
		The finite-order behavior of \(\rho_n(v)\) is shown in Fig.~\ref{comparison} for \(v=0.6\) and \(v=0.956\).
		
		A quantitative comparison with the branch-point/Jacobian extraction, including one additional representative boost, is summarized in Table~\ref{taylor_check}. 
		This provides an independent check that the singularity identified there is indeed the one controlling the convergence radius of the hydrodynamic Taylor series.
		%%%%%%%%%%%%%%%%%%%%%%%%%%%%%%%%%
		\begin{table}[h]
			\centering
			\begin{tabular}{|c|| c c || c|c|}
				\hline\hline
				$v$ &  fit window $[n_{\min},n_{\max}]$  & $k_c^{\rm Taylor}$  & $k_c^{\rm bp/Jac}$  & rel.\ diff. \\
				\hline
				$0.600$   &[60,200]  & \texttt{1.98766}& \texttt{1.98154}& 0.31\% \\
				$0.800$    &[80,200] & \texttt{2.24836} & \texttt{2.23060} & 0.79\% \\
				$0.956$  &[70,500]  & \texttt{3.77659} & \texttt{3.78712} & 0.28\% \\
				\hline\hline
			\end{tabular}
			\caption{Comparison of the convergence radius extracted from the large-order Taylor coefficients with the value obtained from the branch-point/Jacobian construction in the main text.}
			\label{taylor_check}
		\end{table}	
		%%%%%%%%%%%%%%%%%%%%%%%%%%%%%%%%%
		
		More directly, the large-order coefficients are controlled by the singularity of
		\(\omega(k)\) nearest to the origin, regardless of how that singularity is identified. 
		Therefore, if an additional singularity closer than the	Jacobian critical value had been missed, the root estimator would approach that smaller radius. 
		The agreement shown in Table~\ref{taylor_check} provides a direct check that the branch point found from the loss-of-invertibility condition is indeed the convergence-limiting singularity for the representative boost velocities studied.

		\end{document}